\documentclass[trackchanges,twocolumn]{aastex7}
\usepackage{graphicx}
\usepackage{subcaption}
\usepackage{booktabs}
\usepackage{amsmath}
\usepackage{multirow}   
\usepackage{longtable}

\begin{document}

\title{Diversity in Evolutionary Status and Magnetic Activity among Solar-Type Twin Detached Eclipsing Binaries}

\correspondingauthor{Liying Zhu}
\email{zhuly@ynao.ac.cn}

\author[0000-0002-8320-8469]{Fang-Bin Meng}
\affiliation{Yunnan Observatories, Chinese Academy of Sciences, Kunming 650216, People's Republic of China}
\affiliation{University of Chinese Academy of Sciences, No. 1 Yanqihu East Road, Huairou District, Beijing 101408, People's Republic of China}
\email{mengfangbin@ynao.ac.cn}  

\author[0000-0002-0796-7009]{Li-Ying Zhu}
\affiliation{Yunnan Observatories, Chinese Academy of Sciences, Kunming 650216, People's Republic of China}
\affiliation{University of Chinese Academy of Sciences, No. 1 Yanqihu East Road, Huairou District, Beijing 101408, People's Republic of China}
\email{zhuly@ynao.ac.cn}

\author[0000-0002-5995-0794]{Sheng-Bang Qian}
\affiliation{Department of Astronomy, School of Physics and Astronomy, Yunnan University, Kunming 650091, People’s Republic of China}
\email{qiansb@ynu.edu.cn} 

\author[0000-0003-3767-6939]{Lin-Jia Li}
\affiliation{Yunnan Observatories, Chinese Academy of Sciences, Kunming 650216, People's Republic of China}
\email{lipk@ynao.ac.cn} 

\author[0000-0001-5094-3910]{David Mkrtichian}
\affiliation{National Astronomical Research Institute of Thailand, 191 Siriphanich Bldg., Huay Kaew Rd., Chiang Mai 50200, Thailand}
\email{} 

\author[0000-0002-2276-6352]{Nian-Ping Liu}
\affiliation{Yunnan Observatories, Chinese Academy of Sciences, Kunming 650216, People's Republic of China}
\email{lnp@ynao.ac.cn}

\author[0000-0003-1593-1724]{Ahmet Dervişoğlu}
\affiliation{Department of Astronomy and Space Sciences, Atatürk University, Faculty of Science, Erzurum, 25050, Türkiye}
\affiliation{Türkiye National Observatories, DAG, Erzurum, 25050, Türkiye
}
\email{dervisoglu.a@gmail.com} 

\author{Er-Gang Zhao}
\affiliation{Yunnan Observatories, Chinese Academy of Sciences, Kunming 650216, People's Republic of China}
\email{zergang@ynao.ac.cn} 

\author[0000-0001-5511-7183]{Boonrucksar Soonthornthum}
\affiliation{National Astronomical Research Institute of Thailand, 191 Siriphanich Bldg., Huay Kaew Rd., Chiang Mai 50200, Thailand}
\email{boonrucksar@narit.or.th}

\author[0000-0001-7440-9595]{Sergey Zvyagintsev}
\affiliation{Institute of Astronomy, Russian Academy of Sciences, 119017, Pyatnitskaya str., 48, Moscow, Russia}
\email{srgzvgntsv@gmail.com}

\author[]{Somsawat Rattanasoon}
\affiliation{National Astronomical Research Institute of Thailand, 191 Siriphanich Bldg., Huay Kaew Rd., Chiang Mai 50200, Thailand}
\email{sosavar4@gmail.com} 

\author[0000-0003-2017-9151]{Jia Zhang}
\affiliation{Yunnan Observatories, Chinese Academy of Sciences, Kunming 650216, People's Republic of China}
\email{zhangjia@ynao.ac.cn} 

%% Use the \collaboration command to identify collaborations. This command
%% takes an optional argument that is either a number or the word "all"
%% which tells the compiler how many of the authors above the command to
%% show. For example "\collaboration[all]{(DELVE Collaboration)}" wil include
%% all the authors above this command.
%%
%% Mark off the abstract in the ``abstract'' environment. 
\begin{abstract}
We present a combined photometric and spectroscopic analysis of four detached eclipsing binaries (KIC 8957954, KIC 10593759, KIC 8302455, and TIC 207398432), all of which exhibit composite G-type spectra and nearly equal mass ratios. Based on survey data and our own observations, we measured radial velocities with the broadening function method, applied the fd3 program for spectral disentangling, and modeled the light curves with the Wilson–Devinney code to determine accurate absolute parameters. The results reveal significant differences in evolutionary stages and magnetic activity despite their nearly equal masses. Both components of KIC 8957954 and KIC 8302455 are on the main sequence; KIC 10593759 has evolved to the subgiant stage; and in TIC 207398432, the secondary has entered the red giant phase. Stronger magnetic activity is observed in KIC 10593759 and TIC 207398432, characterized by rapid O’Connell Effect Ratio variations, with the latter also exhibiting multiple superflare events. In addition, the spectral characteristics of TIC 207398432 suggest that it may be part of a hierarchical triple system. This study provides precise absolute parameters for twin binaries and offers important observational evidence for understanding their evolutionary diversity, magnetic activity, and the possible presence of tertiary companions.	
	
\end{abstract}

%% Keywords should appear after the \end{abstract} command. 
%% The AAS Journals now uses Unified Astronomy Thesaurus (UAT) concepts:
%% https://astrothesaurus.org
%% You will be asked to selected these concepts during the submission process
%% but this old "keyword" functionality is maintained in case authors want
%% to include these concepts in their preprints.
%%
%% You can use the \uat command to link your UAT concepts back its source.

%% From the front matter, we move on to the body of the paper.
%% Sections are demarcated by \section and \subsection, respectively.
%% Observe the use of the LaTeX \label
%% command after the \subsection to give a symbolic KEY to the
%% subsection for cross-referencing in a \ref command.
%% You can use LaTeX's \ref and \label commands to keep track of
%% cross-references to sections, equations, tables, and figures.
%% That way, if you change the order of any elements, LaTeX will
%% automatically renumber them.
\section{Introduction}
Eclipsing binary systems are fundamental astrophysical laboratories for determining accurate stellar parameters. In particular, double-lined eclipsing binaries (SB2s) allow for precise measurements of stellar masses and radii primarily based on geometric and dynamical modeling, without relying on stellar evolutionary models. The typical uncertainties of these measurements are below 1$\%$ \citep{2010A&ARv..18...67T}, and in some systems even reach 0.2$\%$ \citep{2020MNRAS.498..332M}. Among various types of eclipsing binaries, detached systems are especially valuable because their components evolve nearly independently, without significant mass transfer. As such, they provide ideal testbeds for validating single-star evolutionary models under the influence of a binary environment \citep{2002A&A...396..551L,2017A&A...608A..62H}.

In binary systems, those with mass ratios close to unity are usually referred to as twin binaries. This concept was first introduced by \citet{1979AJ.....84..401L}, who identified a narrow peak in the mass-ratio distribution of SB2s, suggesting that some binaries may preferentially form nearly equal-mass components. Subsequent statistical studies indicated that twin binaries are mainly concentrated in F–G–K spectral types, representing about 3$\%$ of all binaries \citep{2006A&A...457..629L,2009AJ....137.3442S}. However, \citet{2014MNRAS.445.2028C} argued that this fraction may be overestimated due to observational selection effects, since SB2 systems with nearly equal luminosities are more easily recognized as twins. Several formation pathways have been proposed for twin binaries, such as circumbinary disk accretion, disk fragmentation, and multi-body dynamical interactions \citep{2000MNRAS.314...33B,2002MNRAS.336..705B,2000A&A...360..997T}, but decisive evidence is still lacking. Early observational studies primarily relied on spectroscopic statistics, with relatively limited use of light curves (LCs), partly due to the observational challenges of monitoring long-period systems. In recent years, large-scale photometric surveys and space missions, such as the Kepler Space Telescope \citep{2010Sci...327..977B} and the Transiting Exoplanet Survey Satellite (TESS; \citealp{2015JATIS...1a4003R}), have significantly expanded the available data on twin candidates, facilitating LC-based searches. However, cases that combine high-quality LCs with radial velocity (RV) measurements to derive detailed physical properties of twin binaries remain scarce.

Magnetic activity is a common phenomenon among late-type stars, manifested through starspots, flares, chromospheric emission in the Ca II H$\&$K and H$\alpha$ lines, and coronal X-ray radiation \citep{2009A&ARv..17..251S,1996A&A...312..221M,2024ApJ...965..167L}.
Starspots produce out-of-eclipse photometric modulations, which can vary in amplitude and phase, while flares cause rapid, transient brightening \citep{2003A&A...397..285G,2006A&A...446.1129B}. Some solar-like stars are even observed to produce flares with energies several orders of magnitude higher than typical solar events \citep{2013ApJS..209....5S}. 

Against this backdrop, this study analyzes four twin binaries with composite G-type spectra (KIC 8957954, KIC 10593759, KIC 8302455, and TIC 207398432), utilizing data from TESS, Kepler, the Large Sky Area Multi-Object Fiber Spectroscopic Telescope (LAMOST; \citealp{2012RAA....12.1197C}), the Sloan Digital Sky Survey (SDSS; \citealp{2000AJ....120.1579Y}), and our own spectroscopic observations. Despite their nearly equal masses, these systems exhibit marked differences in both evolutionary states and magnetic activity levels. We present detailed analyses of their absolute parameters, evolutionary divergence, and activity-related phenomena.

The paper is organized as follows. In Section \ref{section 2}, we describe all the available observational data. Section \ref{section 3} analyzes the variations in orbital periods. Section \ref{section 4} presents the spectroscopic analysis, while Section \ref{section 5} models the LCs. Finally, Section \ref{section 6} discusses and summarizes the results.

\section{ Observations and Data Reduction}\label{section 2}
\subsection{Photometic Observations}
The Kepler Space Telescope was launched on March 6, 2009. Its field of view is a fixed region of approximately 115 square degrees, located in the direction of the Cygnus–Lyra region. The field is centered roughly at right ascension $19^{\mathrm{h}} 22^{\mathrm{m}} 40^{\mathrm{s}}$ and declination $44^{\circ} 30^{\prime} 00^{\prime \prime}$, situated about $13.5^{\circ}$ above the Galactic plane \citep{2010ApJ...713L..79K}. The Kepler mission operated in two primary sampling modes: Long Cadence with a sampling interval of 29.4 minutes, and Short Cadence with a sampling interval of 58.85 seconds. The instrument has a bandpass spanning 400–900 nm, with peak sensitivity in the 600–750 nm range. For stars with Kepler magnitude kmag $\approx 11$, the photometric precision at 1-minute cadence reaches $\sim$200 parts per million (ppm), approaching the photon noise limit \citep{2010ApJ...713L.160G}.

TESS was launched aboard a SpaceX Falcon 9 rocket in April 2018. TESS is equipped with four wide-field CCD cameras, each with a field of view of $24^{\circ} \times 24^{\circ}$ and an aperture of 10.5 cm. Its bandpass covers approximately 600–1000 nm, with peak sensitivity near 800 nm. For stars with Tmag $\approx 10$, TESS achieves a photometric precision of approximately 200 ppm over a 60-minute integration \citep{2023AN....34430139S}. The mission observes the sky in sequential sectors, each monitored for roughly 27.4 days. The primary data products of TESS include LCs, target pixel files (TPFs), and full-frame images (FFIs). The LCs and TPFs provide high temporal resolution data with 2-minute and 20-second cadences, while the FFIs cover the entire field of view and have improved from an initial cadence of 30 minutes to 10 minutes and 200 seconds during the extended mission.

\begin{figure*}[ht!]
	\centering
	\includegraphics[width=\linewidth]{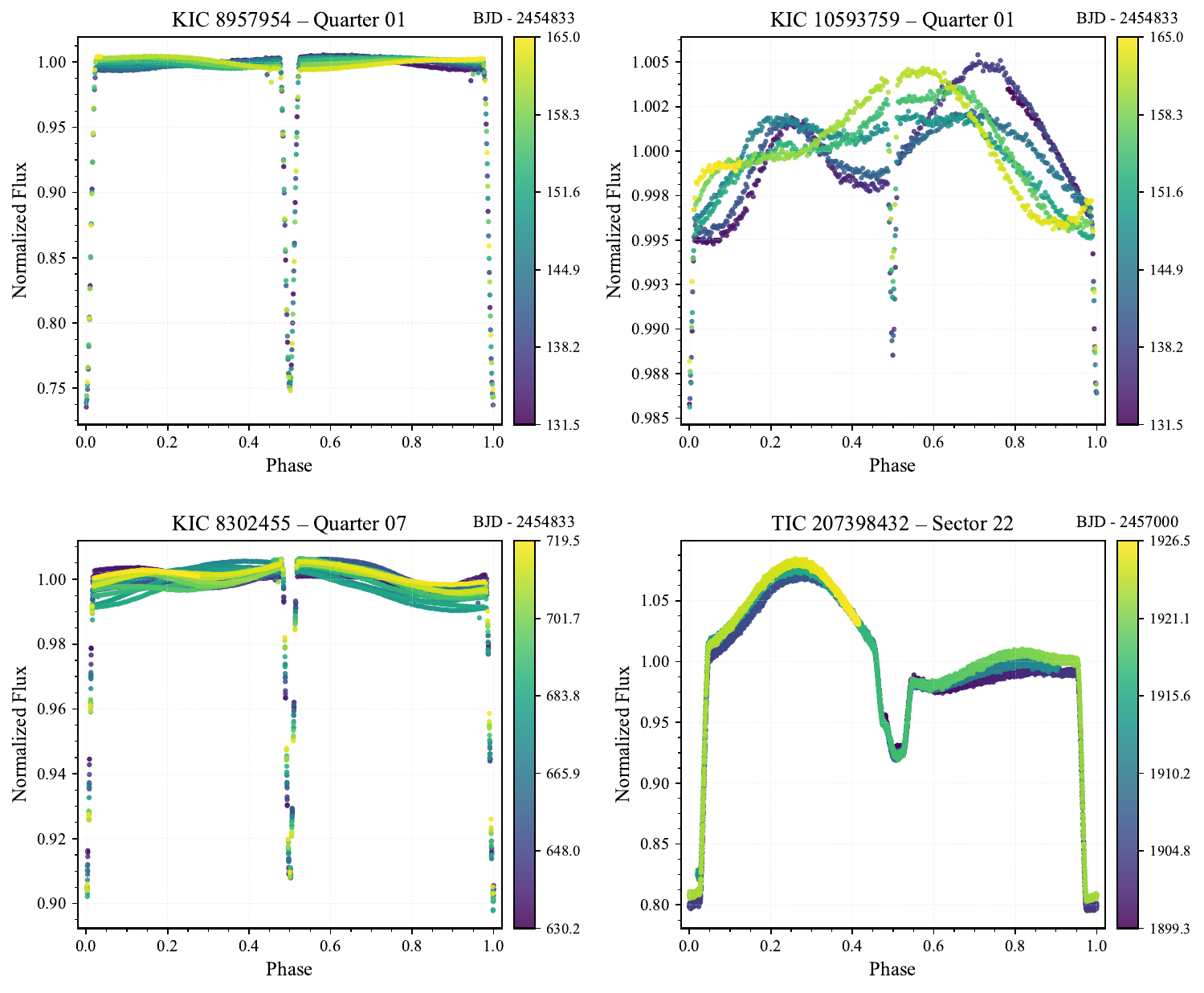}
	\caption{Phase-folded LCs of the four targets, each showing a segment of the data with points color-coded by observation time.}
	\label{Fig.1}
\end{figure*}

We retrieved all available Kepler and TESS photometric data for the four systems from the Mikulski Archive for Space Telescopes (MAST). Both simple aperture photometry (SAP) and pre-search data conditioning–corrected SAP (PDCSAP) LCs are provided by the Science Processing Operations Center pipeline. In our analysis, we adopt the SAP LCs, which preserve the intrinsic eclipse shapes and stellar variability without the additional corrections applied in PDCSAP. For LCs that exhibit obvious long-term instrumental trends, we applied a Locally Weighted Scatterplot Smoothing (LOWESS) to remove these trends while preserving the short-term eclipse features. The resulting detrended LCs were then normalized to their median flux for further analysis. For each system, we show the phase-folded SAP LCs with data points color-coded by observation time, along with one representative continuous segment (approximately one Kepler quarter or one TESS sector; see Figure \ref{Fig.1}).

\subsection{Spectroscopic Observations}
LAMOST, also known as the Guo Shoujing Telescope, is a Schmidt-type reflecting telescope with an effective aperture ranging from 3.6 to 4.9 meters. It is located at the Xinglong Observatory in Hebei Province, China. LAMOST is equipped with a focal plane of 1.75 meters in diameter, capable of accommodating 4000 optical fibers, allowing it to obtain spectra of 4000 celestial objects simultaneously in a single exposure. LAMOST supports two spectral resolution modes. The low-resolution mode (LRS) has a resolving power of $R \approx 1800$ and covers the wavelength range 3700–9000 \AA. The medium-resolution mode (MRS) has $R \approx 7500$ and covers two spectral windows: the blue arm (4950–5350 \AA) and the red arm (6300–6800 \AA) \citep{2015RAA....15.1095L}.  Stellar atmospheric parameters—such as the effective temperature ($T_{\mathrm{eff}}$), surface gravity ($\log g$), and metallicity ($[\mathrm{Fe}/\mathrm{H}]$)—are automatically derived by the LAMOST Stellar Parameter Pipeline \citep{2014IAUS..306..340W}.

SDSS began formal operations in 2000, utilizing a dedicated 2.5-meter telescope located at Apache Point Observatory (APO) in New Mexico, USA. Since the SDSS-IV phase, a second spectroscopic facility has been deployed at Las Campanas Observatory (LCO) in Chile, enabling coordinated observations from both hemispheres. The Apache Point Observatory Galactic Evolution Experiment (APOGEE) was initiated as part of SDSS-III (APOGEE-1), and later expanded in SDSS-IV (APOGEE-2) to extend its coverage across the full sky \citep{2017AJ....154...94M}.  APOGEE conducts spectroscopy in the near-infrared, specifically in the H-band ($1.51–1.70 \mu$m), with a spectral resolving power of approximately $R \sim 22{,}500$.

For our targets, we collected spectra from both LAMOST and SDSS. However, due to the insufficient phase coverage of TIC 207398432, we obtained six additional spectra between January and March 2025 using the 2.4-meter telescope at the Thai National Observatory (TNO), equipped with the Medium Resolution Echelle Spectrograph (MRES). MRES achieves a typical spectral resolution of $R \sim 18{,}000$ over the 380–900 nm wavelength range. The acquired spectra were processed following standard procedures in the Image Reduction and Analysis Facility (IRAF), including extraction and wavelength calibration, and were subsequently normalized to the continuum.

\section{Eclipse Timing Variations and Period Correction}\label{section 3}

Based on the observational data from Kepler and TESS, we used a parabolic fitting method to determine several eclipse timings. The O-C analysis method was then applied to calculate precise phases and study orbital period variations. The periods employed in the initial linear ephemeris for the O–C calculations were taken from the Kepler or TESS eclipsing binary catalogs
 \citep{2016AJ....151...68K,2022ApJS..258...16P}.  In this study, the primary eclipse, defined as phase zero, corresponds to the moment when the more massive component is being eclipsed, as determined from the RV analysis (see Section \ref{section 4.1}).

\subsection{KIC 8957954} 
We computed the times of minima for KIC 8957954 using TESS and Kepler data and constructed the corresponding O–C diagram. During the Kepler observations (2009–2013), the orbital period remained essentially constant, whereas in the TESS observations (2019–2024), low-amplitude variations of approximately 0.004 days were observed. Such small-scale fluctuations are commonly seen in late-type stellar systems and may be related to magnetic activity. No other significant long-term trends are apparent; therefore, we applied a linear ephemeris to correct the period. The ETV curve after subtracting the best-fit linear trend is shown in Figure \ref{Fig.2} (top-left panel). Given the limited temporal coverage, it remains unclear whether these variations are persistent or periodic. The revised linear ephemeris is:

$$
\text{Min.} = \text{BJD}2454967.30922(3) + 4.359849948(61) \times E.
$$

\subsection{KIC 10593759}
For KIC 10593759, the TESS photometry is too noisy to reveal any recognizable eclipse signatures, and therefore no reliable times of minima can be measured from the TESS data. Consequently, only the Kepler  LCs were used to determine the primary and secondary eclipse timings. In the Kepler O-C diagram, the secondary minima show a clear offset relative to the primary ones. This displacement is a purely geometric effect arising from the small but non-zero orbital eccentricity: when $e \cos \omega \neq 0$, the secondary eclipse does not necessarily occur at photometric phase 0.5 , i.e., exactly halfway between two consecutive primary eclipses. The observed separation between the two $\mathrm{O}-\mathrm{C}$ curves therefore reflects the presence of a slight eccentricity in the system.
We thus used the primary minima to refine the orbital period. The O–C variations of the primary and secondary minima after the linear correction are shown in the upper-right panel of Figure \ref{Fig.2}. The resulting linear ephemeris is:

$$
\text{Min.} = \text{BJD}2454959.61524(12) + 6.265209406(924) \times E.
$$

\subsection{KIC 8302455 and TIC 207398432}

\begin{figure*}[htbp]
	\centering
	
	\includegraphics[width=0.48\linewidth]{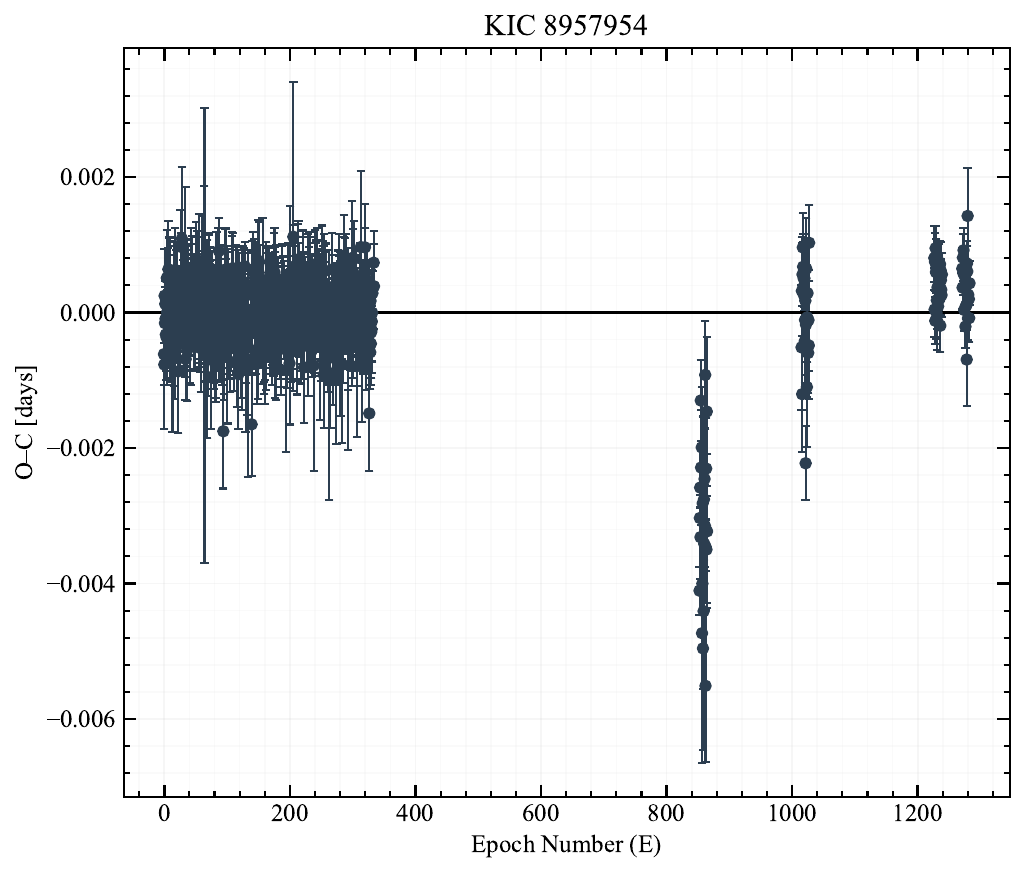}
	\hspace{0em}  % 控制两个图之间的间距
	\includegraphics[width=0.48\linewidth]{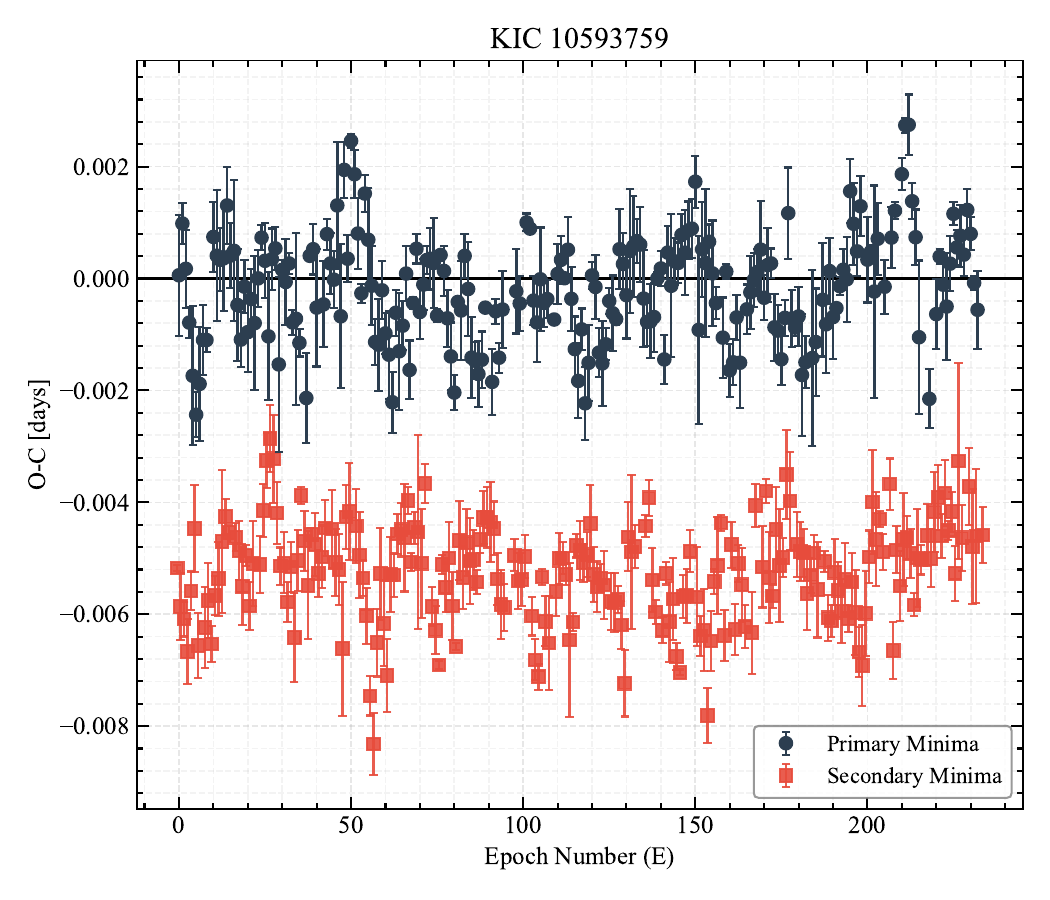}
	
	\vspace{0em}  % 控制上下图之间的间距
	
	\includegraphics[width=0.48\linewidth]{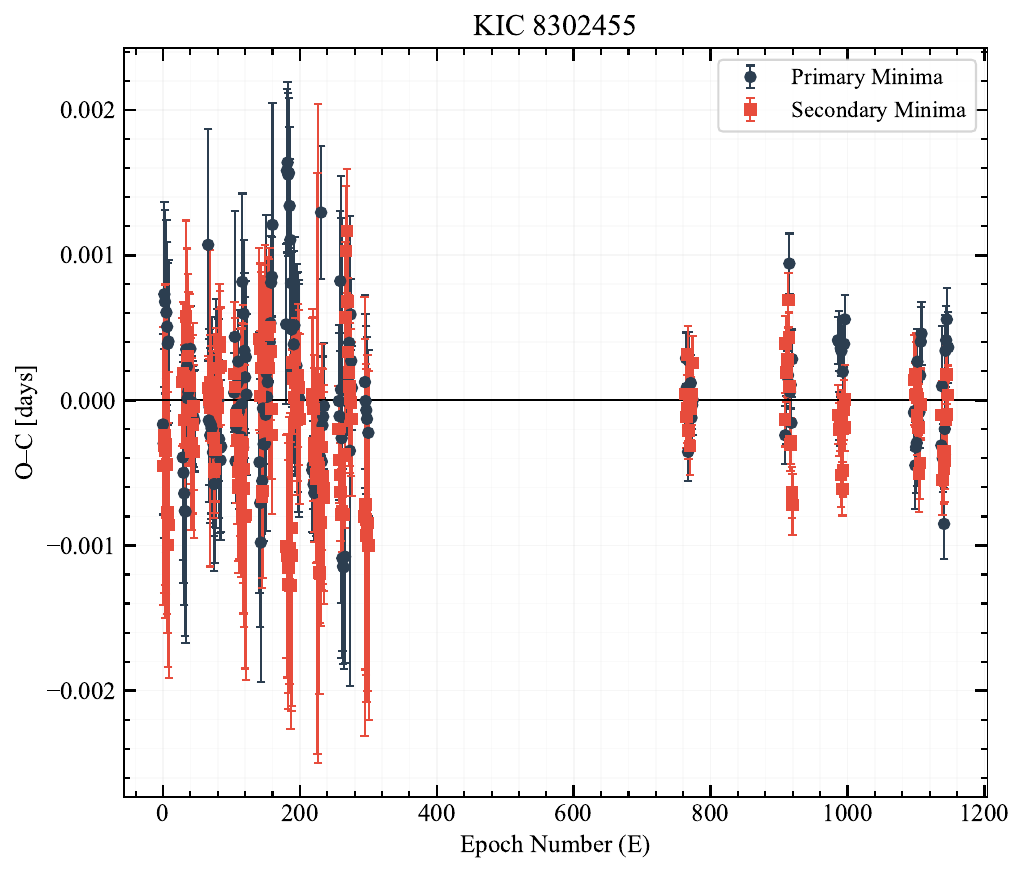}
	\hspace{0em}  % 控制两个图之间的间距
	\includegraphics[width=0.48\linewidth]{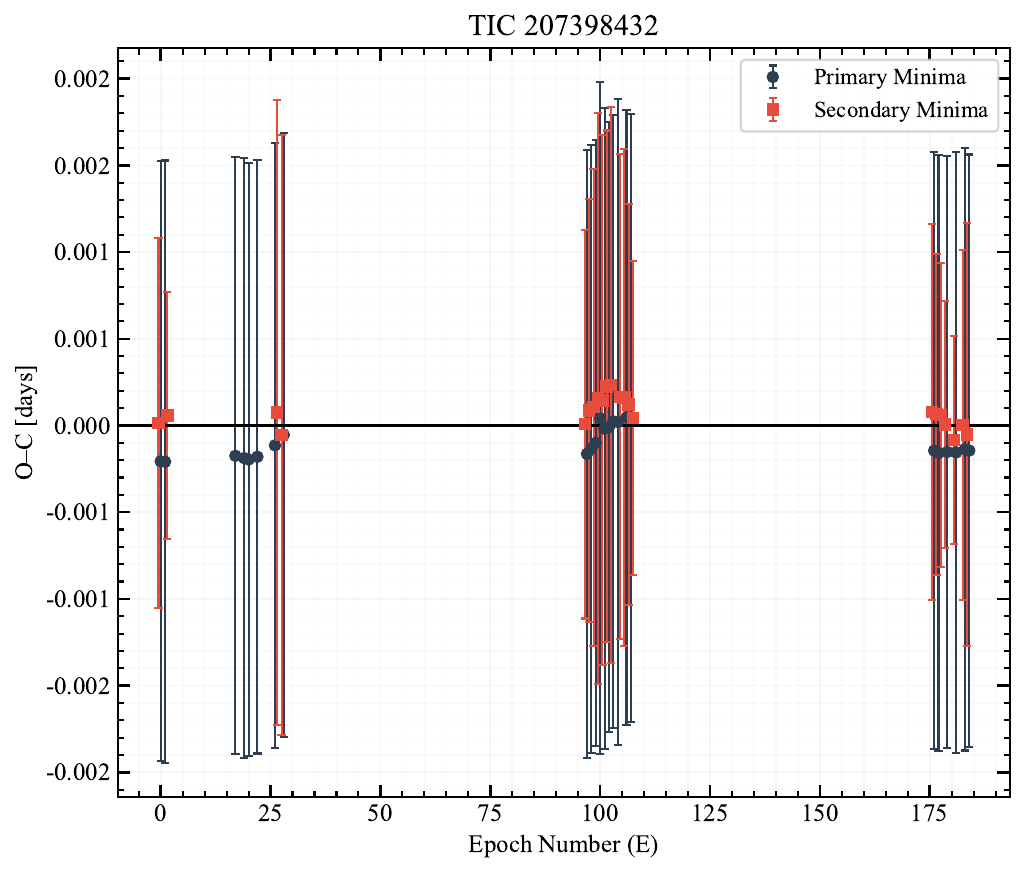}
	
	\caption{The O--C diagrams of the four binaries, constructed using the best-fitting linear ephemerides.}
	\label{Fig.2}
\end{figure*}

For KIC 8302455, the eclipse timings were derived from both the Kepler and TESS LCs, whereas TIC 207398432 is covered only by TESS observations. In the latter system, some secondary-eclipse profiles exhibit noticeable distortions and asymmetric shapes, and the corresponding timings were therefore excluded from the analysis. Neither system shows significant long-term curvature in its O–C diagram, indicating that their orbital periods do not display detectable non-linear evolution over the observational baseline. We therefore adopted simple linear fits to determine the mean periods and used these revised periods to correct all eclipse timings, yielding approximately horizontal O–C curves. The corrected primary and secondary O–C values are displayed in the lower panels of Figure \ref{Fig.2}. It is noteworthy that, in both KIC 8302455 and TIC 207398432, the corrected O–C curves exhibit opposite short-term deviations between primary and secondary minima around zero. Such anti-correlated behavior is frequently observed in close binaries hosting significant stellar activity and is generally attributed to starspot-induced distortions of the eclipse profiles rather than genuine orbital-period variations \citep[e.g.,][]{2013ApJ...774...81T,2021AJ....161...46S,2023AJ....165..247P,2023ApJ...954..111M}. The revised linear ephemerides for the two systems are:

$$
\text{Min.} = \text{BJD}2454955.36421(4) + 4.883976819(46) \times E,
$$
$$
\text{Min.} = \text{BJD}2458746.74324(3) + 9.258174035(241) \times E,
$$ for KIC 8302455 and TIC 207398432, respectively.

\section{Spectroscopic Analysis}\label{section 4}
\subsection{Broadening Functions and Radial Velocities}\label{section 4.1}
\begin{figure*}[htbp]
	\centering
	
	\includegraphics[width=0.32\linewidth]{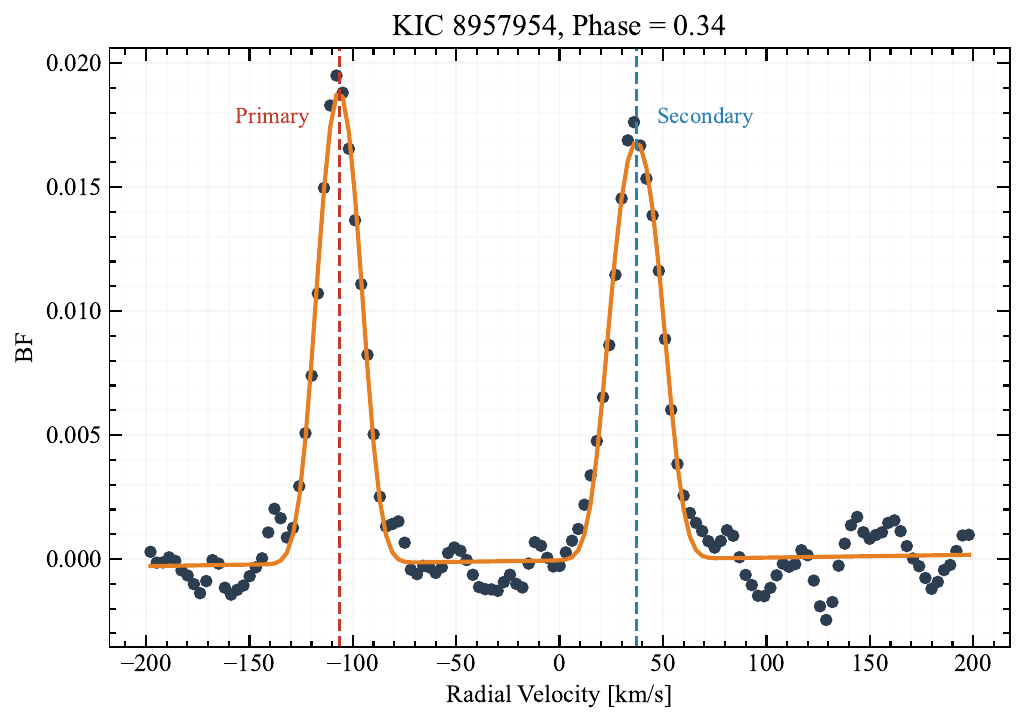}
	\hspace{0em}  % 控制两个图之间的间距
	\includegraphics[width=0.32\linewidth]{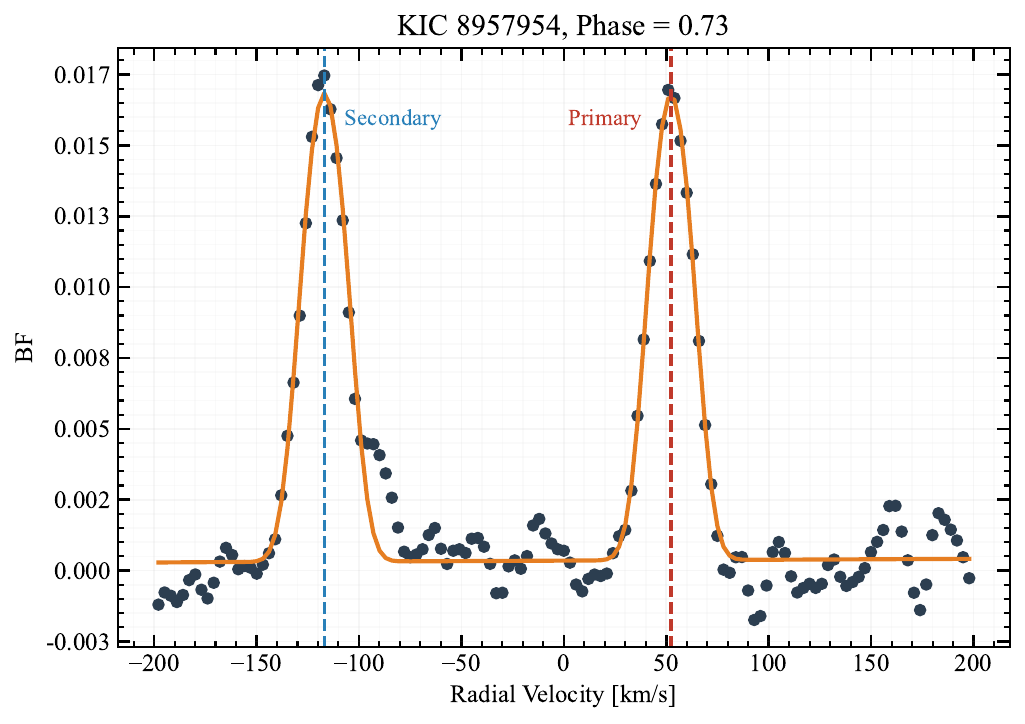}
	\hspace{0em}  % 控制两个图之间的间距
	\includegraphics[width=0.32\linewidth]{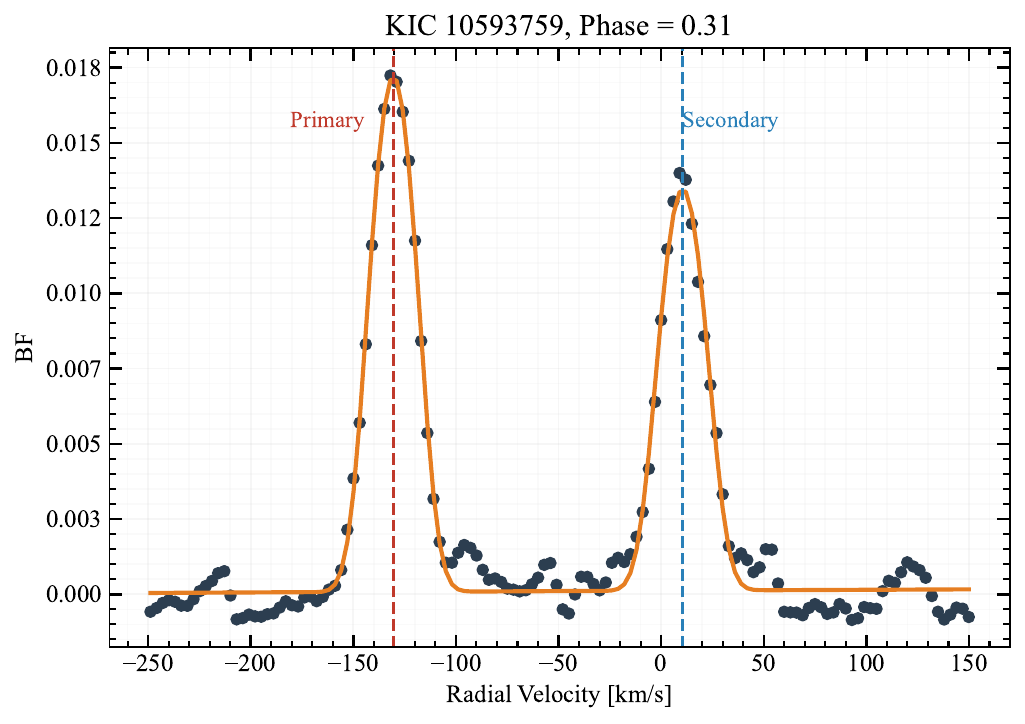}
	
	\vspace{0.5em}
	
	\includegraphics[width=0.32\linewidth]{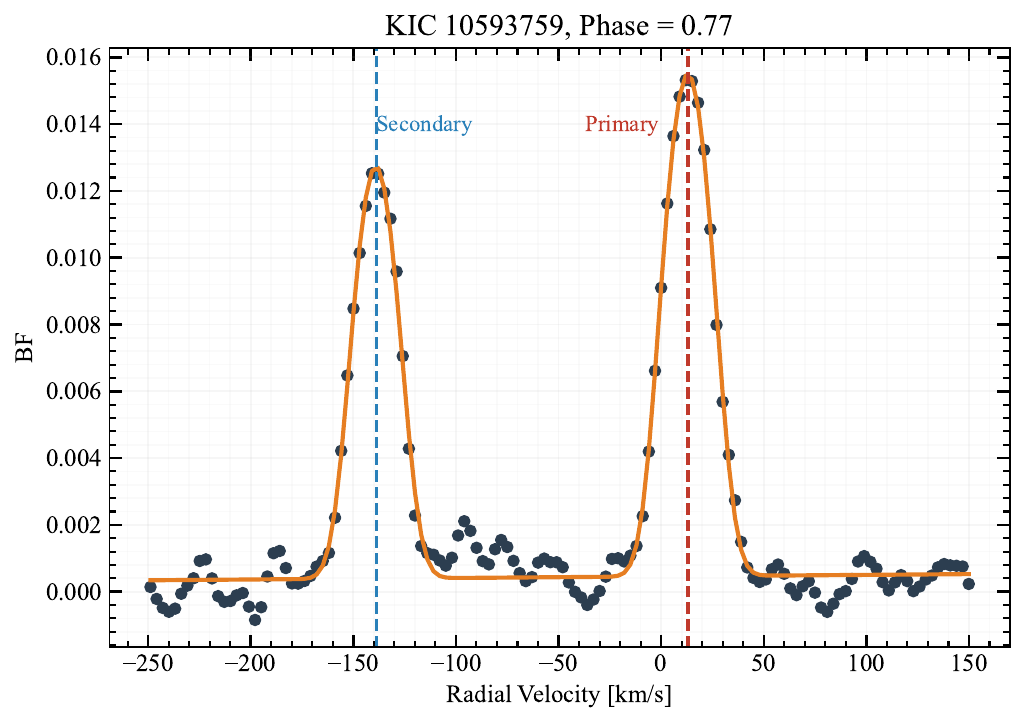}
	\hspace{0em}  % 控制两个图之间的间距
	\includegraphics[width=0.32\linewidth]{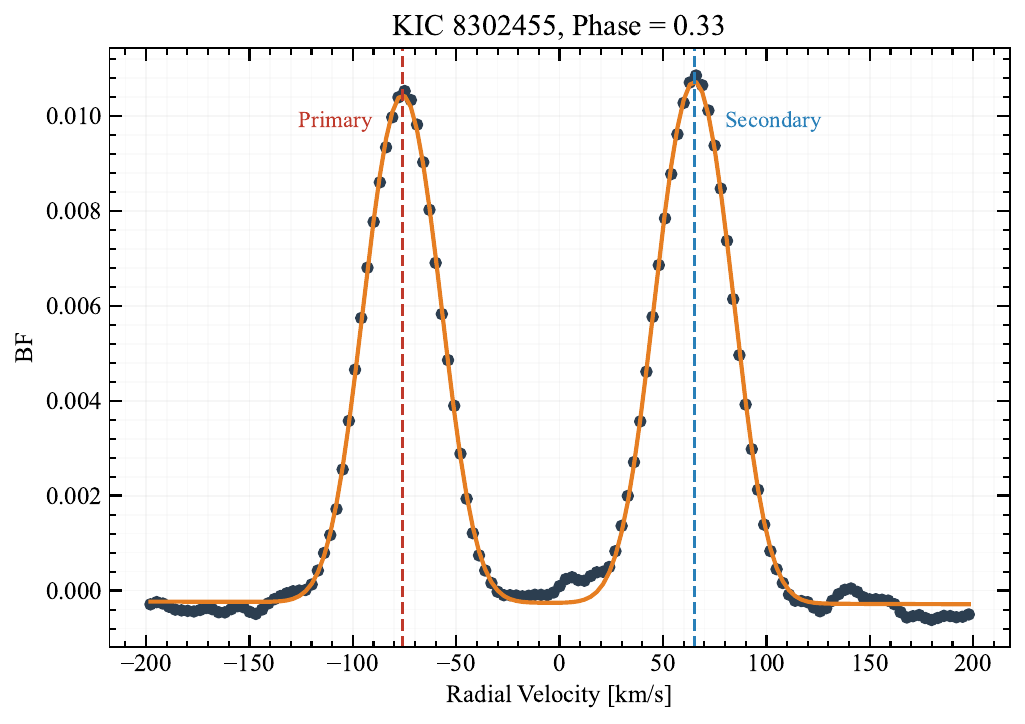}
	\hspace{0em}  % 控制两个图之间的间距
	\includegraphics[width=0.32\linewidth]{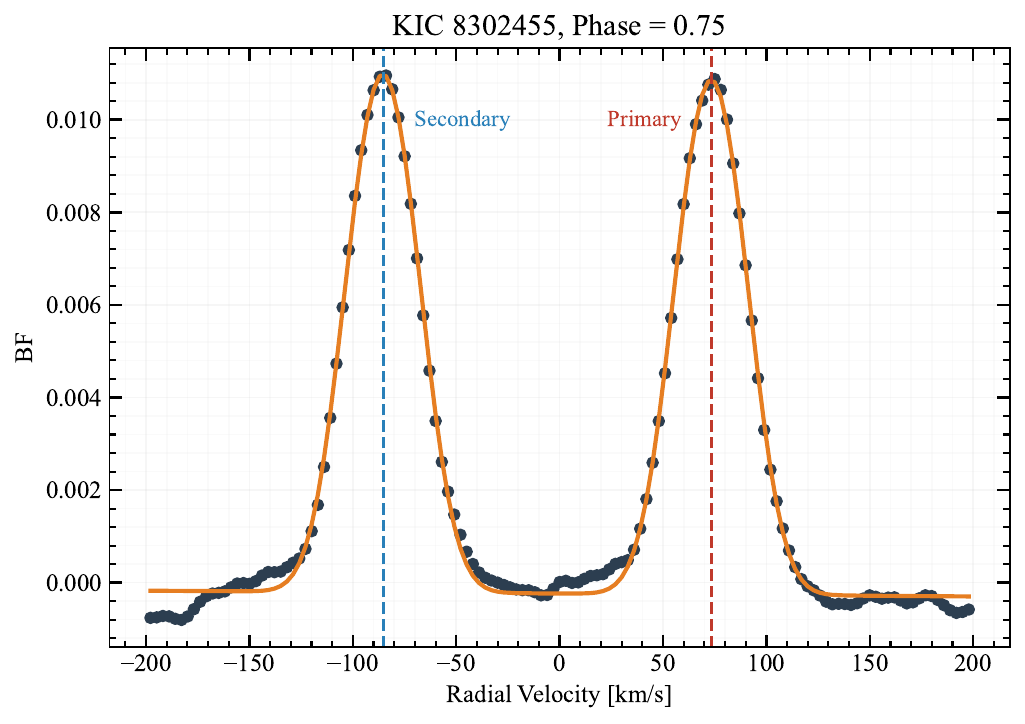}
	
	\vspace{0.5em}
	
	\includegraphics[width=0.32\linewidth]{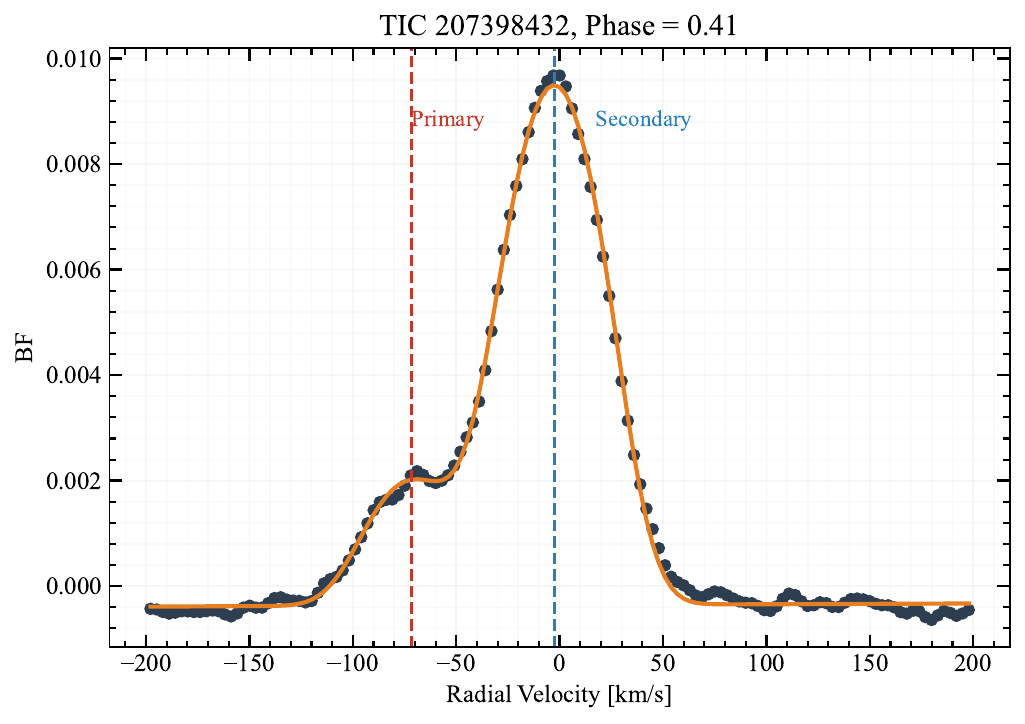}
	\hspace{0em}  % 控制两个图之间的间距
	\includegraphics[width=0.32\linewidth]{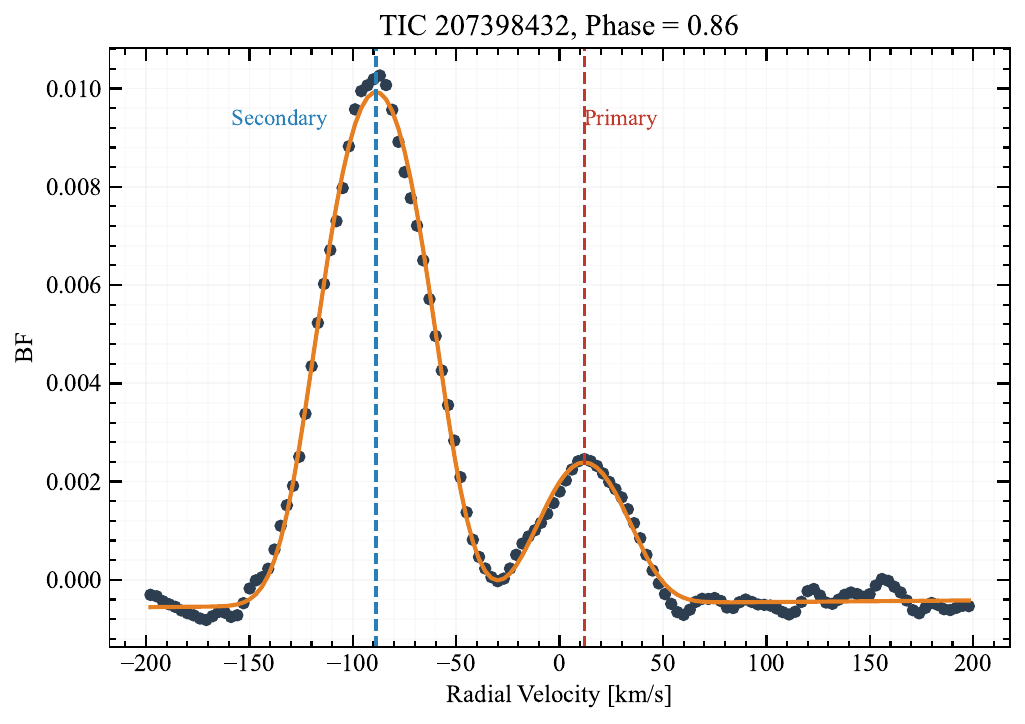}
	\hspace{0em}  % 控制两个图之间的间距
	\includegraphics[width=0.32\linewidth]{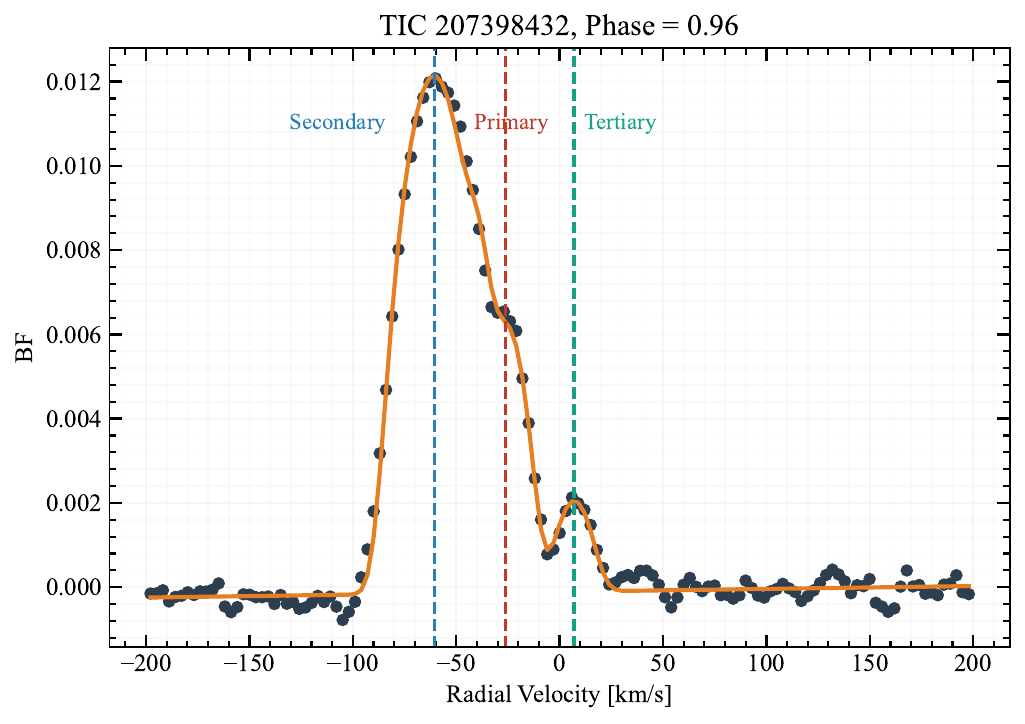}
	
	\caption{The BF profiles of four binaries. The fitted profiles are shown as orange solid lines. The red and blue dashed vertical lines mark the primary and secondary components, respectively. In the final panel, three components are present, with the tertiary component indicated by a green dashed vertical line.}
	\label{Fig.3}
\end{figure*}

We measured the RVs of the targets using the broadening function (BF) technique proposed by \citet{1999TJPh...23..271R}. Template spectra were taken from the PHOENIX synthetic spectral library \citep{2013A&A...553A...6H} and convolved to match the resolution of the observed spectra . The BF profiles encode information from multiple physical effects, including RV, rotational velocity, luminosity ratio, and potential activity regions. Consequently, the BF profiles, together with the eclipse shapes in the LCs, allow a preliminary assessment of the relative evolutionary states of the binary components. Representative BF profiles for the four eclipsing binary systems are shown in Figure \ref{Fig.3}. Most profiles display two prominent peaks, confirming their nature as SB2s, whereas one profile of TIC 207398432 reveals an additional third peak, which may suggest the presence of a tertiary component. The BFs were fitted with Gaussian profiles to derive the RVs, and all measured RVs are listed in Table \ref{tab:rv_long}, with the orbital phases calculated using the linear ephemerides corrected via the O–C analysis in Section \ref{section 3}.

\begin{figure*}[htbp]
	\centering
	
	\includegraphics[width=0.48\linewidth]{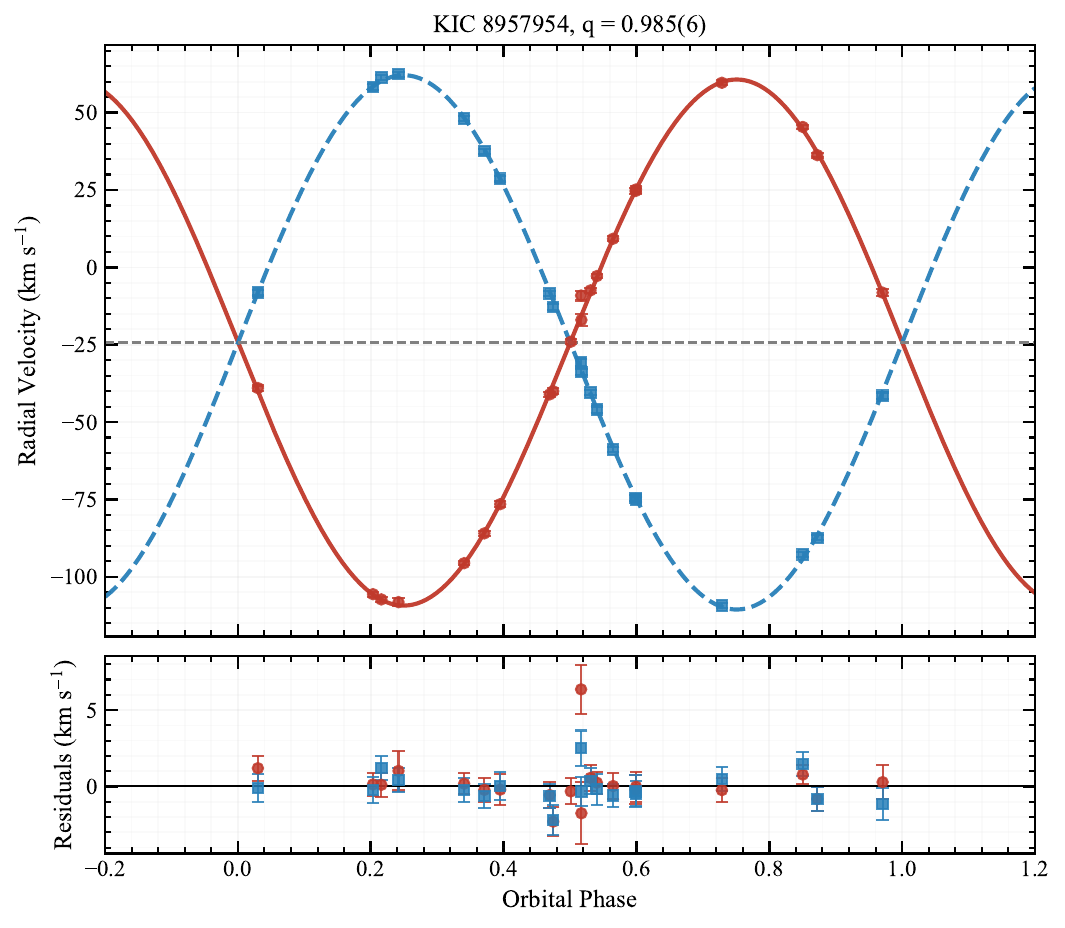}
	\hspace{0em}  % 控制两个图之间的间距
	\includegraphics[width=0.48\linewidth]{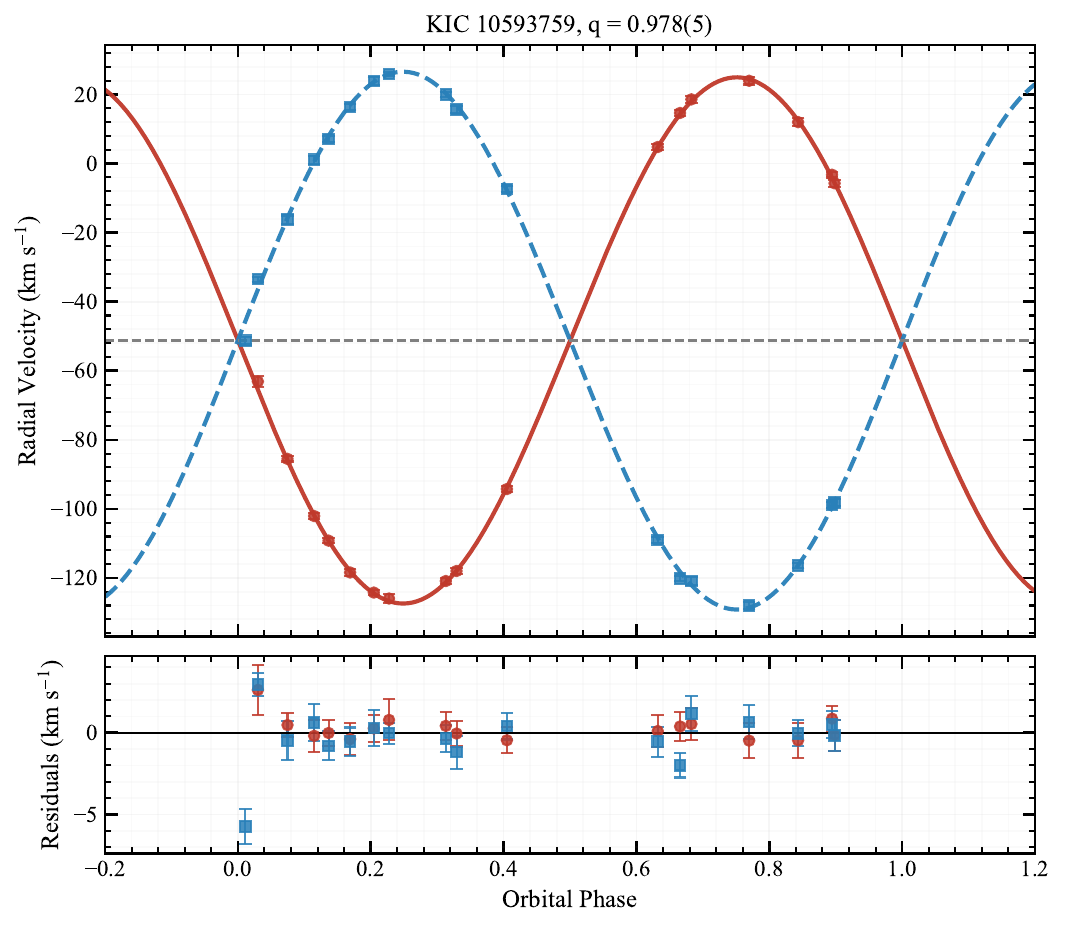}
	
	\vspace{0em}  % 控制上下图之间的间距
	
	\includegraphics[width=0.48\linewidth]{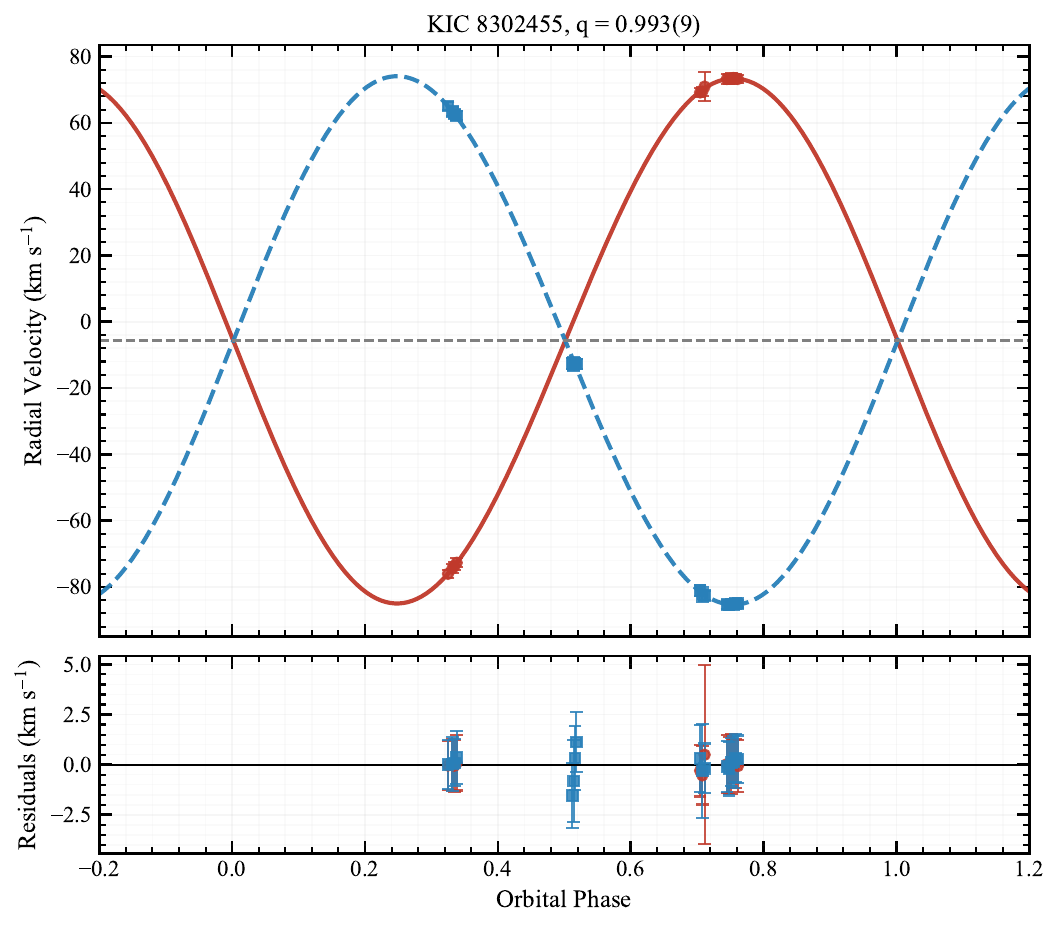}
	\hspace{0em}  % 控制两个图之间的间距
	\includegraphics[width=0.48\linewidth]{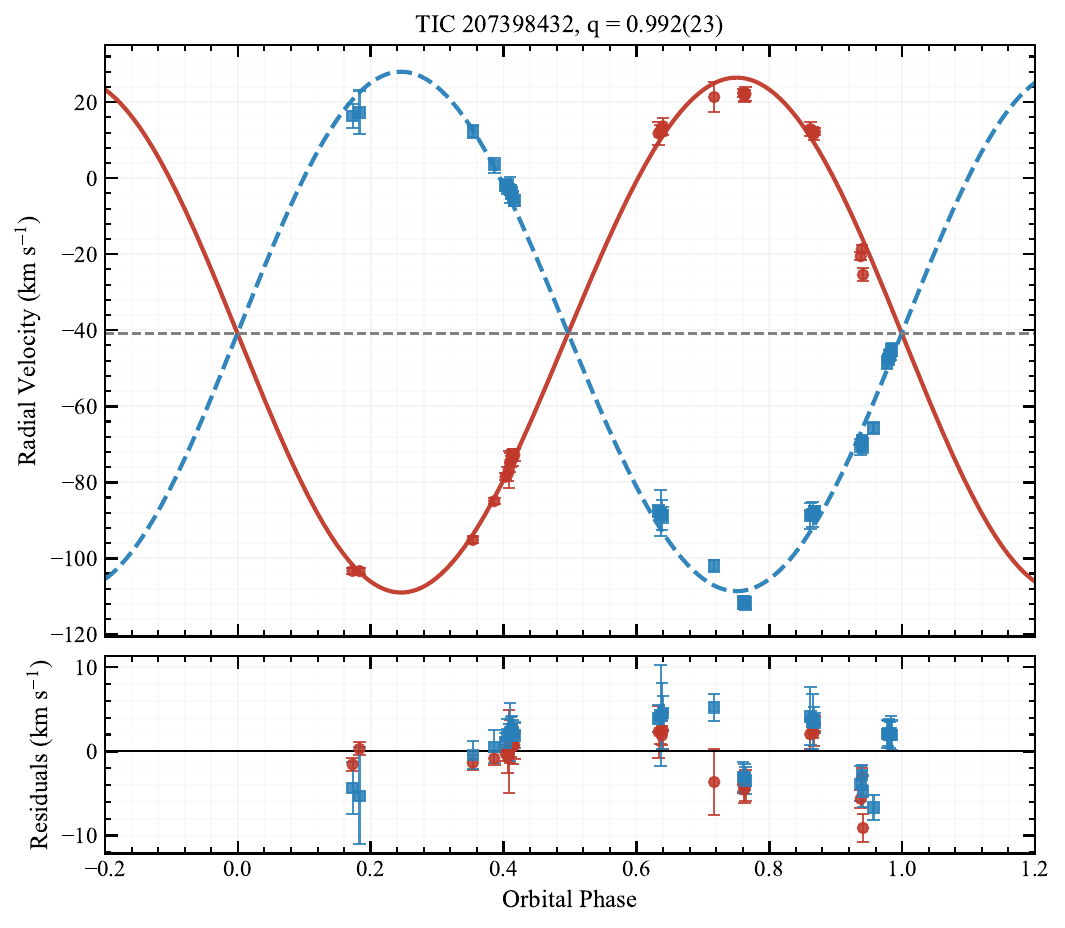}
	
	\caption{The RV curves of the four binaries.}
	\label{Fig.4}
\end{figure*}

The RV curves of the four targets are shown in Figure \ref{Fig.4}. All targets were fitted using a Keplerian orbital model,

\begin{equation}
	V_{\mathrm{rad}}=K[e \cos \omega+\cos (v+\omega)]+\gamma,
\end{equation}

\noindent with the orbital period fixed to that of the corrected linear ephemerides. The model parameters were estimated via nonlinear least-squares fitting, and their uncertainties were derived from a paired bootstrap, resampling the observed RV pairs to preserve the correlation between the two components. The resulting mass ratios are all very close to unity: $q=0.985(6)$ for KIC 8957954, $q=0.978(5)$ for KIC 10593759, $q=0.993(9)$ for KIC 8302455, and $q=0.992(23)$ for TIC 207398432.

\subsection{Spectral Disentangling and Atmospheric Parameters}
\setcounter{footnote}{0}
We performed spectral disentangling analysis for the four binaries using the fd3  code \footnote{\url{http://sail.zpf.fer.hr/fdbinary/}} \citep{2004ASPC..318..111I,2008A&A...482.1031H}. The fd3 program implements the spectral disentangling technique in Fourier space, which does not require any template spectra and can simultaneously determine the orbital parameters and reconstruct the individual component spectra from time-series observations. Since the disentangling procedure requires input light ratio information, we estimated the fractional light contributions of the primary and secondary components from the BF profiles or from the initial LC solution, and adopted these estimates as essential constraints in the disentangling process. In this way, we successfully obtained the individual spectra of the two components for each system.

\begin{figure*}[htbp]
	\centering
	
	\includegraphics[width=1\linewidth]{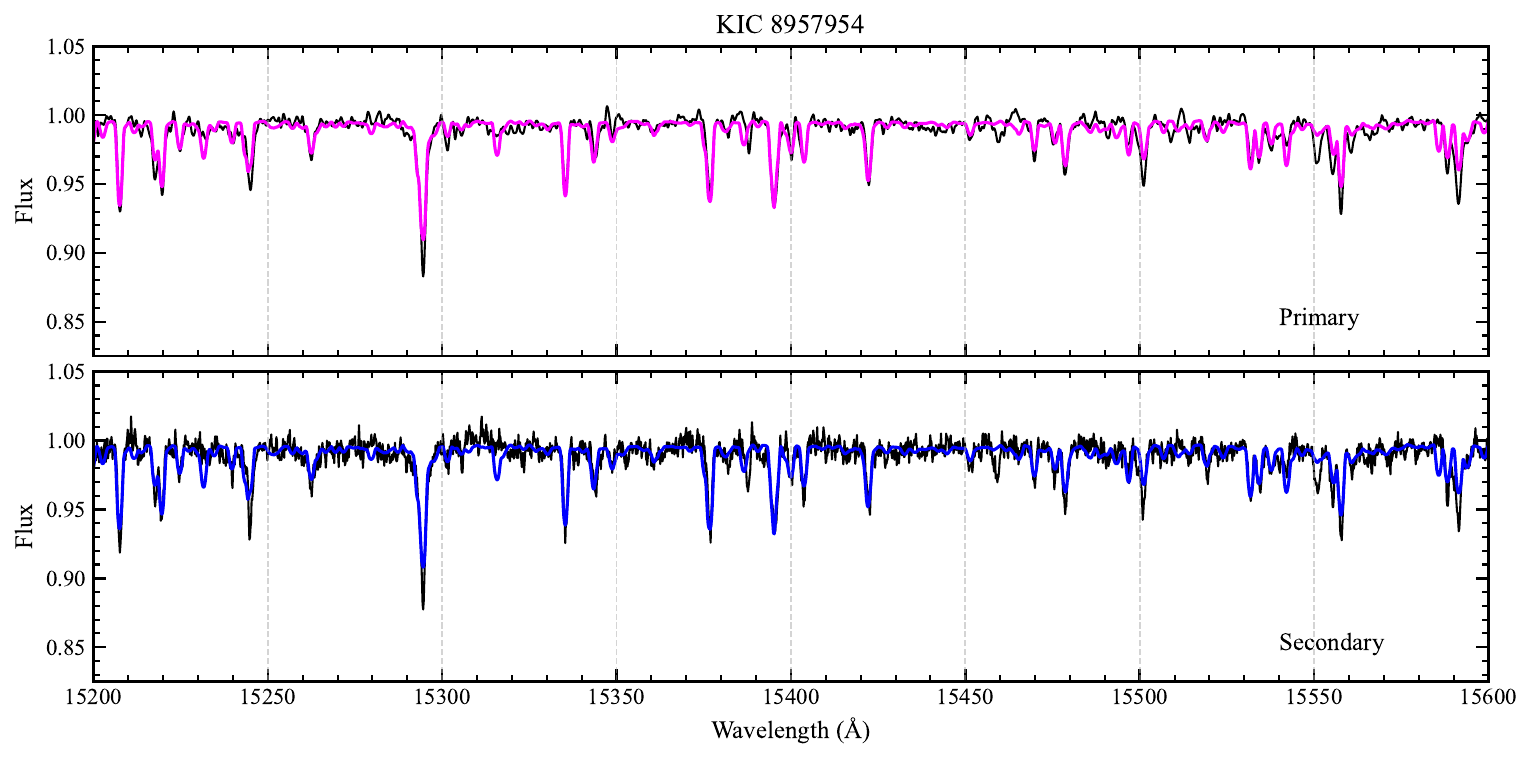}

	\vspace{0em}  % 控制上下图之间的间距
	
	\includegraphics[width=1\linewidth]{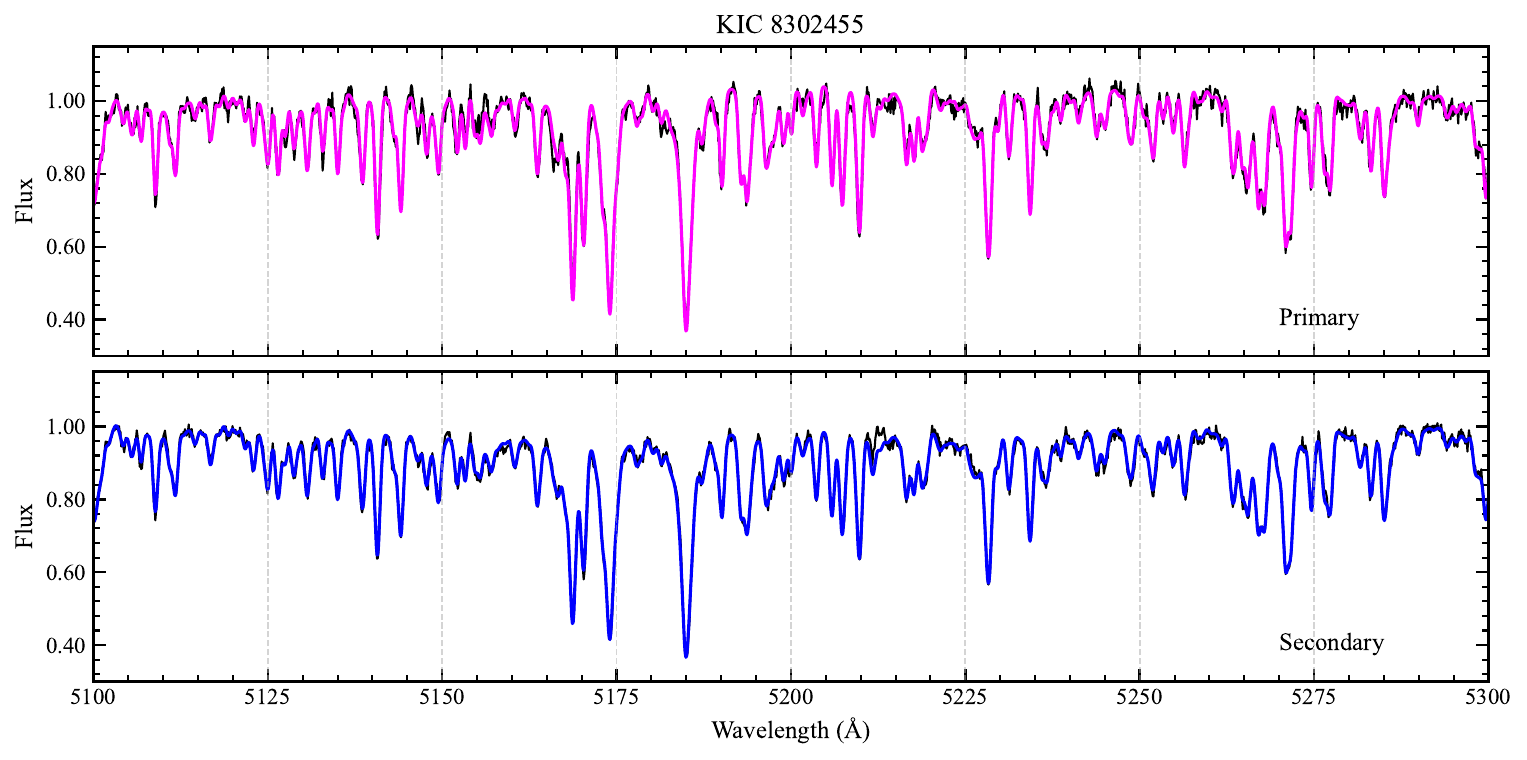}

	\caption{Example disentangled spectra (black) and best-fit synthetic spectra for the primary (pink) and secondary (blue) components of two targets, shown for the SDSS (top panels) and LAMOST (bottom panels) bands.}
	\label{Fig.5}
\end{figure*}

In this study, the results of our spectral disentangling were derived partly from the optical band (LAMOST) and partly from the near-infrared band (SDSS). For the disentangled spectra in the optical band, we employed the University of Lyon Spectroscopic Analysis Software (ULySS; \citealp{2009A&A...501.1269K}) to derive the atmospheric parameters. ULySS is a full-spectrum fitting method that determines the fundamental stellar atmospheric parameters by minimizing the $\chi^2$ between the observed spectrum and the model spectrum.  One of its advantages is that it simultaneously constrains effective temperature, surface gravity, and metallicity during the fitting process, while incorporating a multiplicative polynomial to account for the continuum shape, thereby reducing systematic biases potentially introduced by the pseudo-continuum. We used the ELODIE spectral library  as the template for ULySS to construct the interpolated grid \citep{2001A&A...369.1048P,2007astro.ph..3658P}. Due to the lack of spectral templates suitable for the near-infrared range in ULySS, we used iSpec \citep{2014A&A...569A.111B,2019MNRAS.486.2075B} to determine the atmospheric parameters for the disentangled spectra in the SDSS band. In this process, we simultaneously fitted the disentangled spectra of both components, comparing the synthetic spectra with the disentangled spectra and using the residuals as the convergence criterion. The metallicities of both components were set to be identical, and the remaining parameters were fitted using Markov Chain Monte Carlo (MCMC) to obtain the final results. In Figure \ref{Fig.5}, we show the disentangled spectra and their fits for two representative targets, corresponding to the SDSS and LAMOST bands, respectively. The derived atmospheric parameters are listed in Table \ref{tab:atmparams}, where the iSpec uncertainties appear larger than those from ULySS due to the MCMC-based error estimation adopted in iSpec, rather than any difference in spectral quality.

\begin{deluxetable*}{lcccccc}
	\tablewidth{0pt}  % 自动调整宽度
	\tablecaption{Atmospheric parameters of the four eclipsing binary systems\label{tab:atmparams}}
	\tablehead{
		\colhead{Target} & \colhead{Component} & \colhead{$T_{\rm eff}$ (K)} & \colhead{[Fe/H] (dex)} & \colhead{$\log g$ (cgs)} & \colhead{Method}
	}
	\startdata
	\multirow{2}{*}{KIC 8957954} & Primary   & 5811 $\pm$ 369  & $-0.176 \pm 0.167$ & $4.28 \pm 0.47$ & iSpec \\
	& Secondary & 5780 $\pm$ 402  & $-0.176 \pm 0.167$ & $4.19 \pm 0.44$ & iSpec \\
	\addlinespace
	\multirow{2}{*}{KIC 10593759} & Primary   & 5811 $\pm$ 360  & $-0.169 \pm 0.157$ & $4.48 \pm 0.41$ & iSpec \\
	& Secondary & 5854 $\pm$ 443  & $-0.169 \pm 0.157$ & $4.53 \pm 0.46$ & iSpec \\
	\addlinespace
	\multirow{2}{*}{KIC 8302455} & Primary   & 5931 $\pm$ 20   & $0.053 \pm 0.013$  & $4.34 \pm 0.03$ & ULySS \\
	& Secondary & 5931 $\pm$ 14   & $0.017 \pm 0.009$  & $4.39 \pm 0.02$ & ULySS \\
	\addlinespace
	\multirow{2}{*}{TIC 207398432} & Primary   & 6184 $\pm$ 41   & $-0.392 \pm 0.027$ & $3.83 \pm 0.05$ & ULySS \\
	& Secondary & 4781 $\pm$ 6    & $-0.525 \pm 0.006$ & $2.82 \pm 0.02$ & ULySS \\
	\enddata
\end{deluxetable*}

\section{Light Curve Modeling}\label{section 5}
To derive the system parameters and absolute parameters of the four eclipsing binaries, we modeled their LCs using the Wilson–Devinney (W–D; \citealp{1971ApJ...166..605W,2014ApJ...780..151W}) code with the detached configuration (Mode 2). The mass ratios were fixed to the values obtained from the RV solutions in Section \ref{section 4.1}. Although the W-D code allows for a simultaneous fit of LCs and RVs, we adopted the strategy of determining the mass ratio from RVs alone and then modeling the LCs independently. This approach not only simplifies the LC modeling but also avoids the risk that combined LC+RV solutions may prioritize convergence at the expense of photometric details (e.g., spot-induced modulations). For validation, we also tested joint LC+RV fits and confirmed that the resulting mass ratios are consistent with those from the RV-only solutions within the uncertainties. Since all four targets are late-type systems, the gravity-darkening and bolometric albedo parameters were fixed at $g_1 = g_2 = 0.32$ and $A_1 = A_2 = 0.5$, respectively. The limb-darkening effect was treated with the logarithmic law ($ld = -2$), and the corresponding coefficients were determined by interpolation from the tables of \citet{1993AJ....106.2096V}. In the LC fitting, we fixed the temperature of the primary component to the value obtained from the disentangled spectra (see table \ref{tab:atmparams}), while allowing the secondary temperature to vary.

\subsection{KIC 8957954}\label{section 5.1}
The presence of starspots introduces significant out-of-eclipse distortions in the LC. However, in eclipsing binaries, key geometric parameters—such as the radius ratio—are primarily constrained by the morphology of the eclipses themselves, particularly the shape of ingress and egress and the contact phases. For KIC 8957954, the relatively high orbital inclination produces deep and well-defined eclipses, while the contribution from ellipsoidal variations is modest. Under these conditions, the eclipse morphology alone provides a reliable constraint on the system's geometric parameters.

Based on this consideration, we removed the out-of-eclipse variations using the LOWESS method and fitted the resulting LC. This procedure preserves the eclipse morphology while minimizing the influence of starspots (e.g., \citealt{2022AJ....163..235L}). To verify that this treatment does not introduce systematic bias, we also fitted a segment of the original, uncorrected LC exhibiting relatively weak out-of-eclipse variations (BJD 2455276–2455371). The two approaches yield consistent results (see Figure~\ref{Fig.6}).

Because the two components in this system have nearly identical radii, with a radius ratio $R_2/R_1$ close to unity, the LC does not strictly constrain the radius ratio: the same eclipse morphology can correspond to $R_2/R_1>1$ or $R_2/R_1<1$, representing a local degeneracy. This degeneracy arises from the weak coupling between parameters: the LC mainly constrains the contact phases and the ingress/egress shapes, so swapping nearly equal radii can produce nearly identical residuals.

The uncertainties provided by the W-D code are based on a local linearization around the optimal solution and therefore tend to underestimate the true uncertainties in the presence of parameter coupling and degeneracy. To obtain more realistic uncertainties, we adopted a systematic $\Omega_1$ search for all four systems. The primary surface potential $\Omega_1$ was first fixed and scanned over a physically reasonable range, retaining all convergent solutions and identifying the minimum-residual solution and the $\Omega_1$ boundaries. The solutions at the minimum and at both boundaries were then re-optimized with $\Omega_1$ treated as a free parameter. The fully relaxed minimum-residual solution was adopted as the final result, and the deviations of the boundary solutions from it were used to define the parameter uncertainties. The final parameters are listed in Table~\ref{tab:WD}.

\subsection{KIC 10593759}
KIC 10593759 exhibits the largest photometric variations with an extremely short timescale, showing significant changes in almost every orbital cycle. This introduces substantial challenges for LC modeling. Unlike KIC 8957954, its relatively low orbital inclination results in shallow eclipses and prominent ellipsoidal variations, making it unsuitable to constrain geometric parameters by fitting eclipse profiles after smoothing the out-of-eclipse modulation.

To address this, we directly modeled the observed LC, incorporating multiple starspots to account for the out-of-eclipse variations. Given the long exposure time (1800 s) and the limited number of data points per orbit, we reconstructed each orbital cycle by multiplying the de-trended LC from a single Kepler Quarter with the corresponding out-of-eclipse modulation, effectively increasing the number of points and improving the fit. We performed the fit on a segment spanning BJD 2454965–2454972. Both the orbital eccentricity and argument of periastron were allowed to vary, following indications from the O–C analysis. The resulting model LC is shown in the upper-right panel of Figure \ref{Fig.6}.

\subsection{KIC 8302455}
For KIC 8302455, we selected the segment spanning BJD 2455129-2455154 for the fitting, with an exposure time of 60 s. During this interval, the LC morphology remained nearly unchanged. A pronounced decrease in flux is seen near phase 0.5, corresponding to the occultation of the secondary component. To reproduce this feature, we introduced two starspots on the primary. We also fitted the de-trended LC with out-of-eclipse variations removed (as described in Section \ref{section 5.1}), yielding results consistent with the spotted model.

\subsection{TIC 207398432}
From the BF profiles (see Figure \ref{Fig.3}), the primary of TIC 207398432 appears less luminous than the secondary. The LC shows a pronounced difference in eclipse depths between the two components, indicating a temperature contrast. Moreover, the flat-bottomed feature at phase 0.0 suggests that the primary is fully eclipsed by the secondary. Taken together, these characteristics imply that the secondary is more evolved than the primary. Accordingly, in the fitting process, we adopted a larger surface potential for the primary relative to that of the secondary.

\begin{figure*}[htbp]
	\centering
	
	\includegraphics[width=0.48\linewidth]{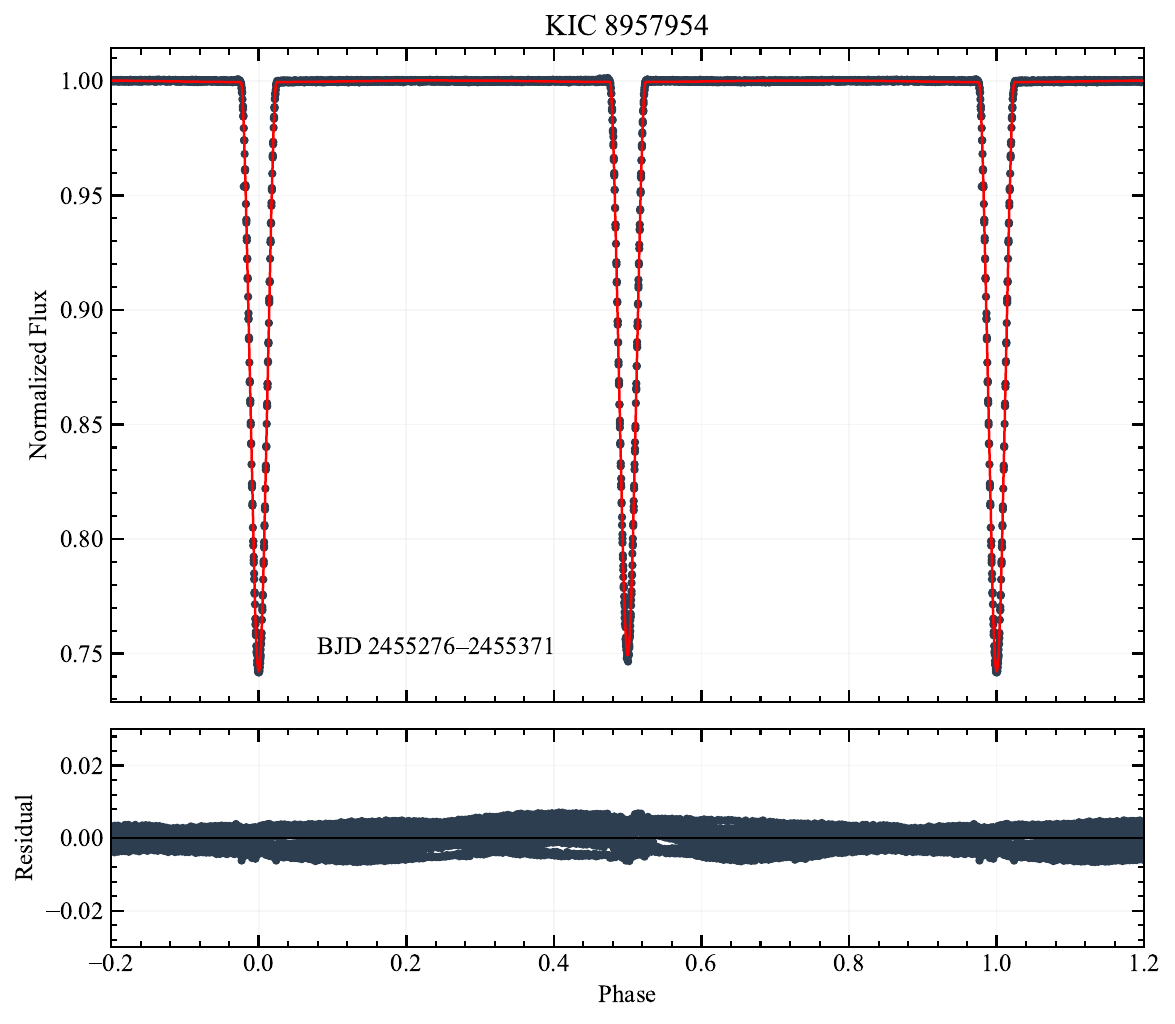}
	\hspace{0em}  % 控制两个图之间的间距
	\includegraphics[width=0.48\linewidth]{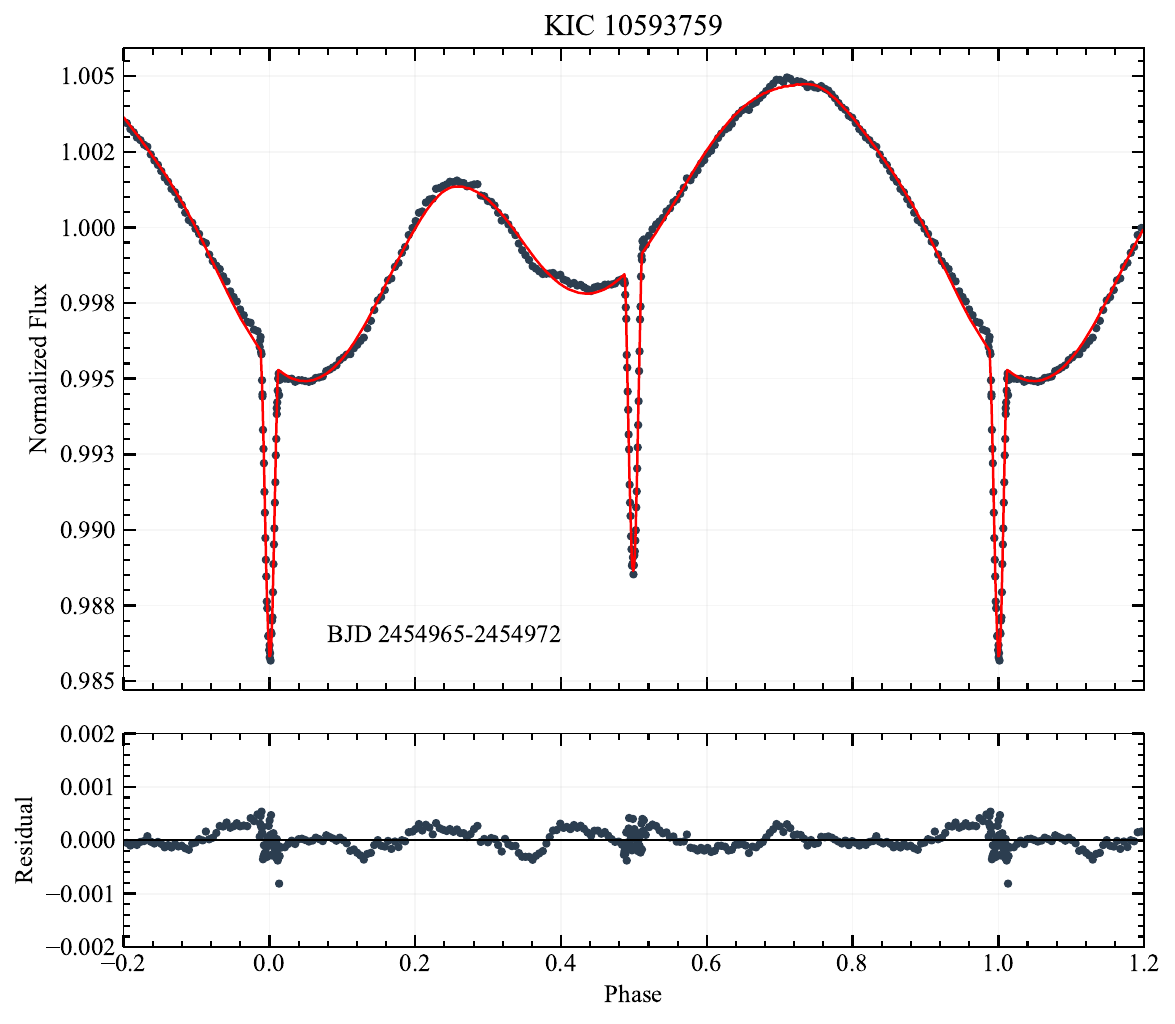}
	
	\vspace{0em}  % 控制上下图之间的间距
	
	\includegraphics[width=0.48\linewidth]{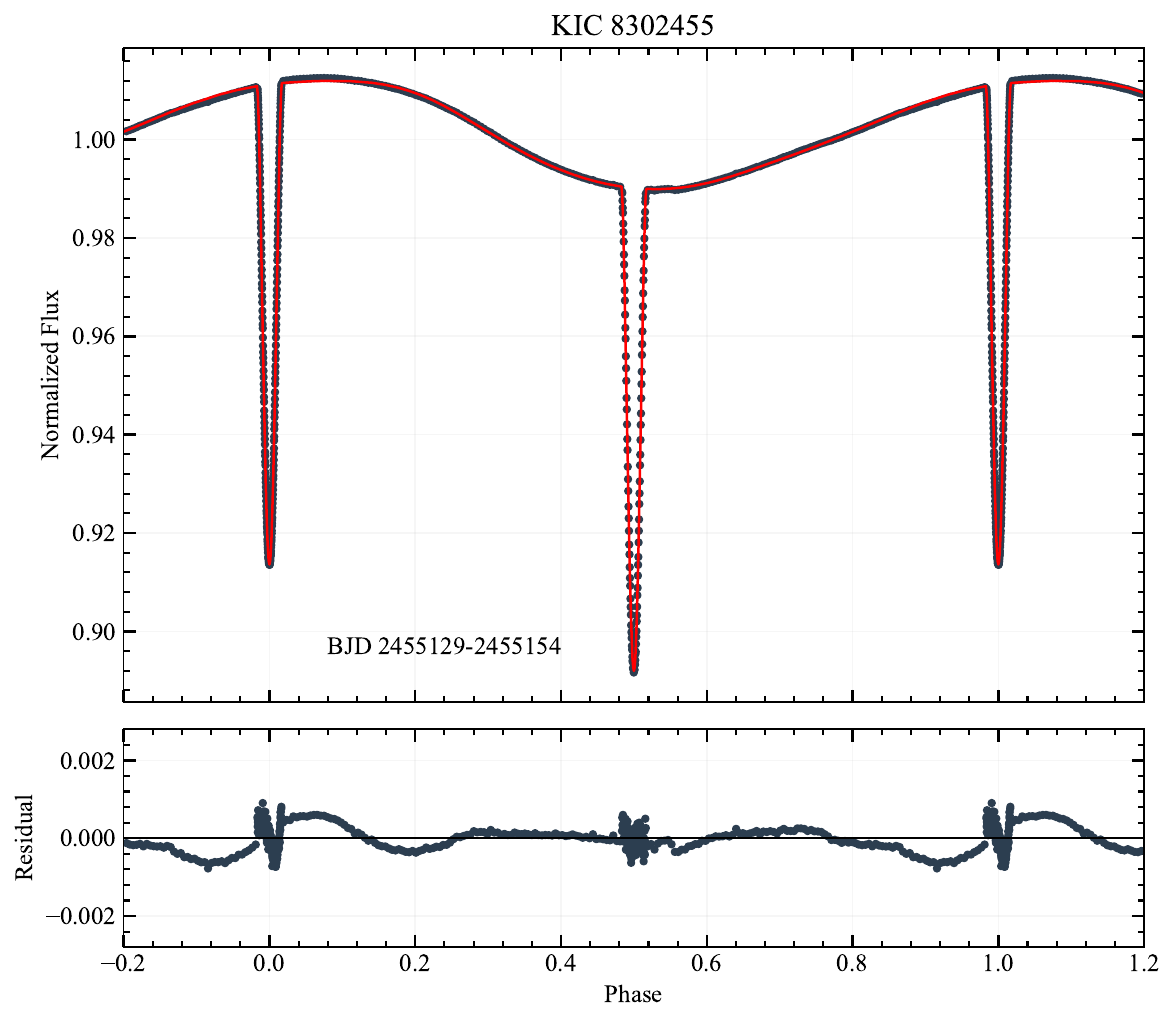}
	\hspace{0em}  % 控制两个图之间的间距
	\includegraphics[width=0.48\linewidth]{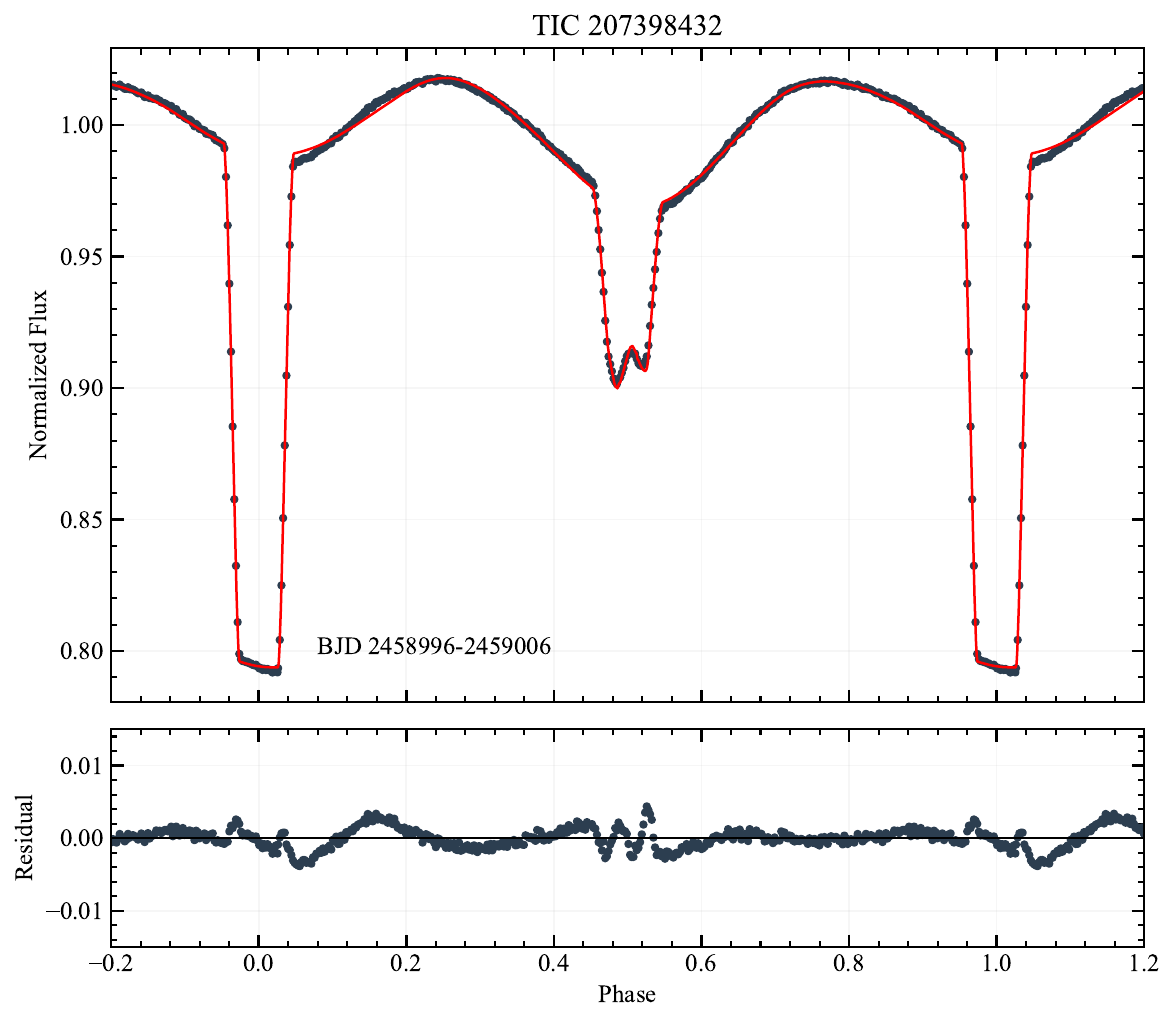}
	
	\caption{Observed (black dots) and best-fit (red lines) LCs of the four binaries. Spot models were included when necessary to reproduce out-of-eclipse modulations. }
	\label{Fig.6}
\end{figure*}

The LC exhibits a prominent and persistent brightening near phase 0.5 (see Figures \ref{Fig.1} and \ref{Fig.6}, bottom-right panels), which occurs when the secondary is occulted by the primary. This feature can be naturally explained by a long-lived cool spot located on the hemisphere of the secondary facing the primary: when the spot is hidden during the secondary eclipse, the overall system brightness temporarily increases, producing the observed hump; when the spot becomes visible, the flux slightly decreases on both sides of the eclipse. The phase stability and longevity of this feature suggest that the spot may be associated with a tidally preferred longitude of magnetic activity. Tidal forcing can modify convective flows and magnetic field topology in close binaries (e.g., \citet{2024A&A...690A.201W}), which may in turn promote the formation of spots near the line connecting the two components.

To account for the irregular variations in the LC, we introduced a starspot model. Fits to data from different sectors yielded consistent geometric parameters (e.g., relative radii, surface potentials, and fill-out factors), while the derived secondary temperature showed slight variations (by a few hundred kelvin), likely affected by the presence of starspots. The fitting result for one LC segment (BJD 2458996–2459006) is shown in the bottom-right panel of Figure \ref{Fig.6}.

\begin{deluxetable*}{lcccc}
	\tablewidth{0pt}
	\tablecaption{Orbital Solutions for the four eclipsing binary systems\label{tab:WD}}
	\tablehead{
		\colhead{star} & \colhead{KIC 8957954} & \colhead{KIC 10593759} & \colhead{KIC 8302455} & \colhead{TIC 207398432}
	}
	\startdata
	Mode	&	Detached Binary	&	Detached Binary	&	Detached Binary	&	Detached Binary	\\
	g1	&	0.32	&	0.32	&	0.32	&	0.32	\\
	g2	&	0.32	&	0.32	&	0.32	&	0.32	\\
	A1	&	0.5	&	0.5	&	0.5	&	0.5	\\
	A2	&	0.5	&	0.5	&	0.5	&	0.5	\\
	$i$(deg)	&	88.79(38)	&	80.19(47)	&	84.30(4)	&	88.79(38)	\\
	$e$	&	$\ldots$	&	0.0015(3)	&	$\ldots$	&	$\ldots$	\\
	$\omega$(rad)	&	$\ldots$	&	2.63(39)	&	$\ldots$	&	$\ldots$	\\
	$T_1$(K)	&	5811(fixed)	&	5811(fixed)	&	5931(fixed)	&	6184(fixed)	\\
	$T_2/T_1$	&	0.994(2)	&	1.002(6)	&	1.000(7)	&	0.770(29)	\\
	$q(M_2/M_1)$	&	0.985(fixed)	&	0.978(fixed)	&	0.993(fixed)	&	0.992(fixed)	\\
	$\Omega_1$	&	12.94(1.11)	&	11.71(90)	&	14.85(97)	&	16.81(36)	\\
	$\Omega_2$	&	13.44(93)	&	11.67(25)	&	14.76(69)	&	5.37(2)	\\
	$L_1/(L_1+L_2)_{\text{band}}$	&	0.54(8)	&	0.51(3)	&	0.50(6)	&	0.17(2)	\\
	$L_2/(L_1+L_2)_{\text{band}}$	&	0.47(8)	&	0.49(3)	&	0.50(6)	&	0.83(2)	\\
	$R_2/R_1$	&	0.95(11)	&	0.98(8)	&	1.00(11)	&	3.63(8)	\\
	$r_{1 \text { pole }}$	&	0.084(7)	&	0.094(7)	&	0.072(5)	&	0.063(1)	\\
	$r_{1 \text { point }}$	&	0.084(7)	&	0.094(7)	&	0.072(5)	&	0.063(1)	\\
	$r_{1 \text { side }}$	&	0.084(7)	&	0.094(7)	&	0.072(5)	&	0.063(1)	\\
	$r_{1 \text { back }}$	&	0.084(7)	&	0.094(7)	&	0.072(5)	&	0.063(1)	\\
	$r_{2 \text { pole }}$	&	0.079(7)	&	0.092(2)	&	0.072(3)	&	0.226(1)	\\
	$r_{2 \text { point }}$	&	0.079(7)	&	0.092(2)	&	0.072(3)	&	0.235(1)	\\
	$r_{2 \text { side }}$	&	0.079(7)	&	0.092(2)	&	0.072(3)	&	0.229(1)	\\
	$r_{2 \text { back }}$	&	0.079(7)	&	0.092(2)	&	0.072(3)	&	0.233(1)	\\
	$f_{fill,1}$	&	1.1(2)$\%$	&	1.5(3)$\%$	&	0.7(2)$\%$	&	0.5(1)$\%$	\\
	$f_{fill,2}$	&	0.9(2)$\%$	&	1.4(1)$\%$	&	0.7(1)$\%$	&	22.3(2)$\%$	\\
	\tableline
	\multicolumn{5}{c}{Absolute parameters} \\
	\tableline
	$M_1(M_\odot)$	&	1.16(1)	&	1.26(1)	&	1.04(2)	&	1.22(4)	\\
	$M_2(M_\odot)$	&	1.14(1)	&	1.23(1)	&	1.03(1)	&	1.21(4)	\\
	$R_1(R_\odot)$	&	1.24(10)	&	1.82(14)	&	1.11(8)	&	1.58(4)	\\
	$R_2(R_\odot)$	&	1.17(10)	&	1.78(4)	&	1.11(5)	&	5.72(6)	\\
	$L_1(L_\odot)$	&	1.57(47)	&	3.38(98)	&	1.38(21)	&	3.26(18)	\\
	$L_2(L_\odot)$	&	1.37(41)	&	3.26(83)	&	1.37(14)	&	15.06(2.29)	\\
	$\log g_1(cgs)$	&	4.32(7)	&	4.02(7)	&	4.36(7)	&	4.13(2)	\\
	$\log g_2(cgs)$	&	4.36(7)	&	4.03(2)	&	4.36(4)	&	3.00(6)	\\
	$a(R_\odot)$	&	14.81(3)	&	19.38(6)	&	15.43(7)	&	24.92(23)	\\
	\enddata
\end{deluxetable*}

\section{Discussions and Conclusion}\label{section 6}
\subsection{Triple System Candidate}
In the spectral analysis of TIC 207398432, one of the TNO spectra exhibits a BF profile with three distinct peaks, suggesting the possible presence of a third body. The signal is significantly above the noise level and thus appears credible. However, no comparable features are detected in the other spectra. The spectrum showing the triple-peak structure corresponds to orbital phase 0.96, which may indicate that at this phase the Doppler shifts of the binary components and that of the putative third body are sufficiently separated to allow its detection. A search of spectra obtained at similar phases (available only from LAMOST) did not reveal a comparable signal. Several explanations can be considered: the spectral signature of the third body may generally be blended with those of the binary components; the spectral resolution of the LAMOST data may be insufficient to resolve the feature; or the RV of the third body may have varied, causing its signal to become undetectable at other epochs. In summary, while the triple-peak BF profile provides tentative evidence for a third body in TIC 207398432, further high-quality observations will be required to confirm its presence.

\subsection{evolution stage}
\begin{figure*}[htbp]
	\centering
	
	\includegraphics[width=1\linewidth]{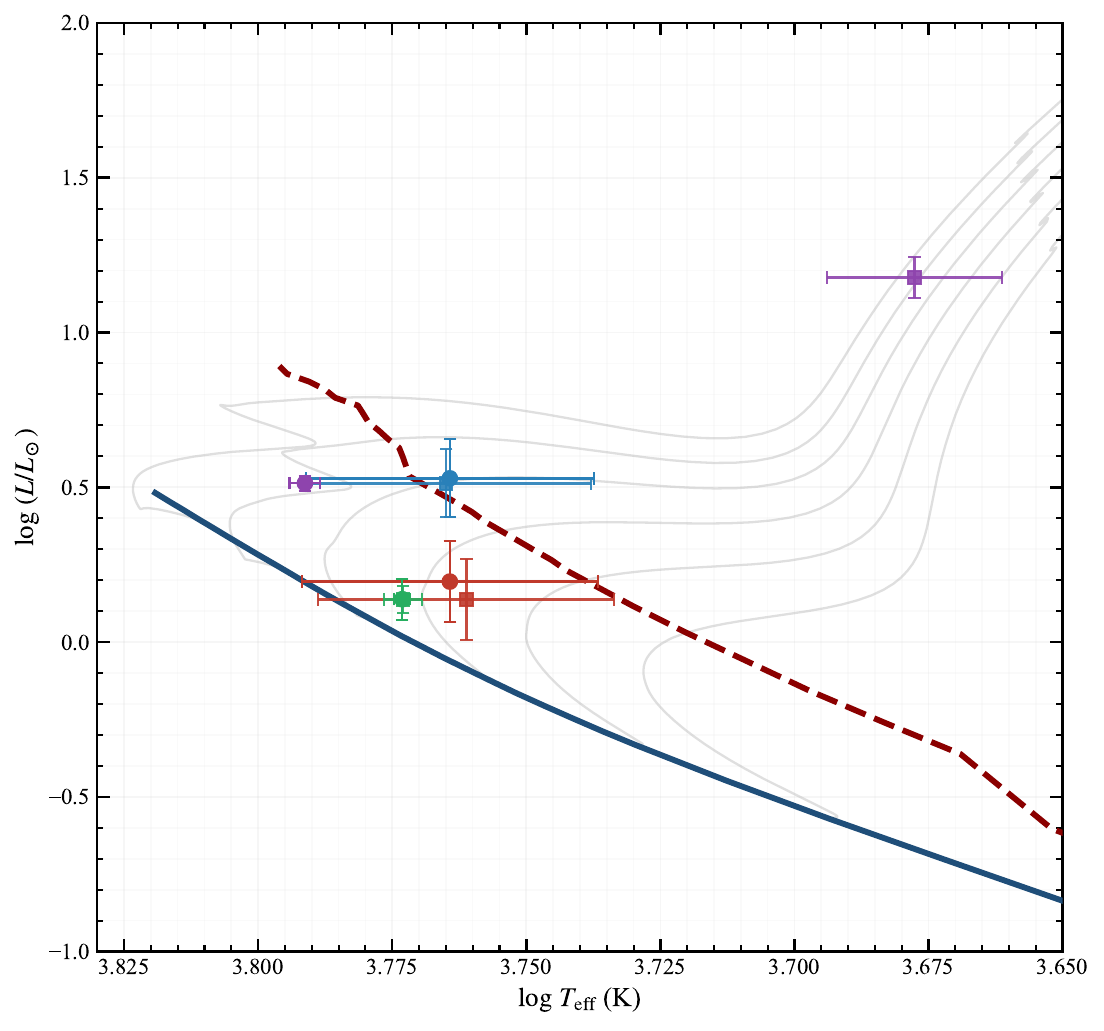}

	\caption{Hertzsprung–Russell diagram showing the evolutionary status of the four analysed twin binary systems. Grey curves represent stellar evolutionary tracks with initial masses between 0.8 and 1.3~$M_\odot$. The solid dark-blue and dashed dark-red lines indicate the zero-age main sequence (ZAMS) and terminal-age main sequence (TAMS), respectively. Circles and squares denote the primary and secondary components, respectively. Colours correspond to different systems: red—KIC~8957954, blue—KIC~10593759, green—KIC~8302455, and purple—TIC~207398432.}
	\label{Fig.7}
\end{figure*}

 Based on the mass functions, RV solutions, and photometric modeling, we derived the physical parameters of the four binary systems, which are summarized in the lower part of Table~\ref{tab:WD}. To better characterize their evolutionary status, we further analyzed classical evolutionary indicators, such as their positions in the Hertzsprung–Russell diagram. The evolutionary tracks were taken from the MESA Isochrones and Stellar Tracks (MIST; \citealp{2016ApJS..222....8D,2016ApJ...823..102C}). As shown in Figure~\ref{Fig.7}, the evolutionary states of the four systems can be broadly divided into three categories:

(1) KIC 8957954 and KIC 8302455 are located on the main sequence, with their two components evolving in nearly synchronous tracks. 

(2) In KIC 10593759, both components are located  near the terminal-age main sequence and are 
	likely evolving off the main sequence into the subgiant phase.

(3) TIC 207398432 exhibits a pronounced difference in evolutionary state between its two components: one star is located near the terminal-age main sequence, while the other has already evolved onto the red-giant branch. Although the RV analysis yields nearly identical masses for the two components, the associated uncertainties allow for the possibility that the more evolved star is slightly more massive. We applied an isochrone-based grid-search method following \citet{2019ApJS..244...43Z}, simultaneously constraining the observed masses, radii, effective temperatures, and metallicity of both components. The search was performed within $\pm3\sigma$ of the observational uncertainties for each parameter. We find that, within these uncertainties, both stars can be consistently reproduced by a single isochrone with a common age of $\log{Age} \approx 9.7$. Within this framework, a coeval binary in which the initially slightly more massive component has evolved onto the red-giant branch while its companion remains near the end of the main sequence can be naturally explained by standard stellar evolutionary tracks. This analysis is based on single-star models and does not include binary interaction effects. Therefore, obtaining the full picture of TIC 207398432's evolutionary history will require detailed binary evolution modeling in the future.

Overall, although the component stars in these four binaries have nearly equal masses, they occupy distinctly different evolutionary stages. This diversity provides useful empirical constraints for studies of stellar evolution, particularly in twin binary systems.

\subsection{magnetic avtivity}
\begin{figure*}[htbp]
	\centering
	
	\includegraphics[width=1\linewidth]{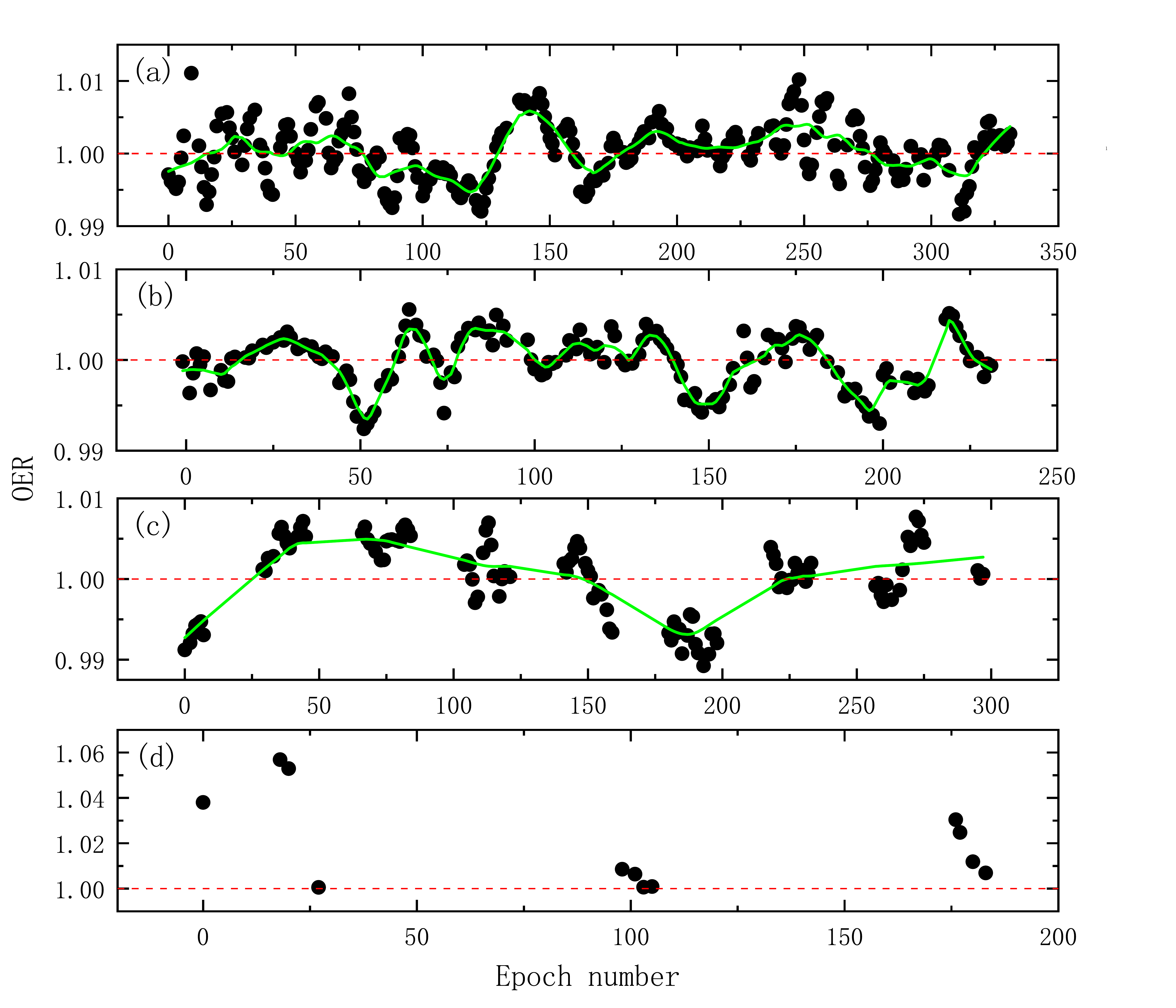}

	\caption{Variations of the O’Connell Effect Ratio (OER) with epoch for the four systems: (a) KIC 8957954, (b) KIC 10593759, (c) KIC 8302455, and (d) TIC 207398432.}
	\label{Fig.8}
\end{figure*}

We further investigated the manifestations of magnetic activity in these four systems. First, for the asymmetry in the LCs caused by starspots, we adopted the O’Connell Effect Ratio (OER) as a quantitative diagnostic \citep{1999PhDT........38M,2022ApJS..262...10K}, defined as

\begin{equation}
	\mathrm{OER}=\frac{\sum_{i=1}^{n / 2} I_i-I_0}{\sum_{i=(n / 2)+1}^n I_i-I_0}
\end{equation}

\noindent  where $I_i$ denotes the flux at a given phase point and $I_0$ is the baseline flux. By comparing the integrated fluxes of the two half-cycles, OER effectively characterizes the degree of LC asymmetry. In this work, the OER is calculated using out-of-eclipse data only, with the baseline flux $I_0$ defined as the mean out-of-eclipse flux for each orbital cycle. This approach significantly reduces the impact of eclipse geometry and system-dependent normalization, which otherwise can strongly affect OER amplitudes. We computed the temporal variations of OER for the four systems at different epochs, with the starting point of the epochs consistent with that used in Section~\ref{section 3}. To present the overall trend more intuitively, we smoothed the results, shown as the green curve in Figure~\ref{Fig.8}. Although no clear periodicity can be identified, the amplitude and variation rate of OER fluctuations provide useful diagnostics of the relative strength of spot-induced light-curve asymmetry and its temporal evolution.

In terms of amplitude, TIC 207398432 exhibits the largest OER variations, while KIC 10593759, KIC 8957954, and KIC 8302455 show comparable fluctuation amplitudes. This indicates that, among the four systems, the spot-induced asymmetry is most pronounced in TIC 207398432, whereas the other three systems display similar levels of asymmetry. Regarding the variation rate, KIC 10593759 shows the fastest OER fluctuations, followed by KIC 8957954 and then KIC 8302455, suggesting progressively slower evolution of the dominant spot-induced asymmetry patterns. For TIC 207398432, the limited temporal coverage prevents a robust assessment of its long-term variability, although a rapid decrease in OER amplitude is observed between $E=22$ and $E=25$.

\begin{figure*}[htbp]
	\centering
	
	\includegraphics[width=1\linewidth]{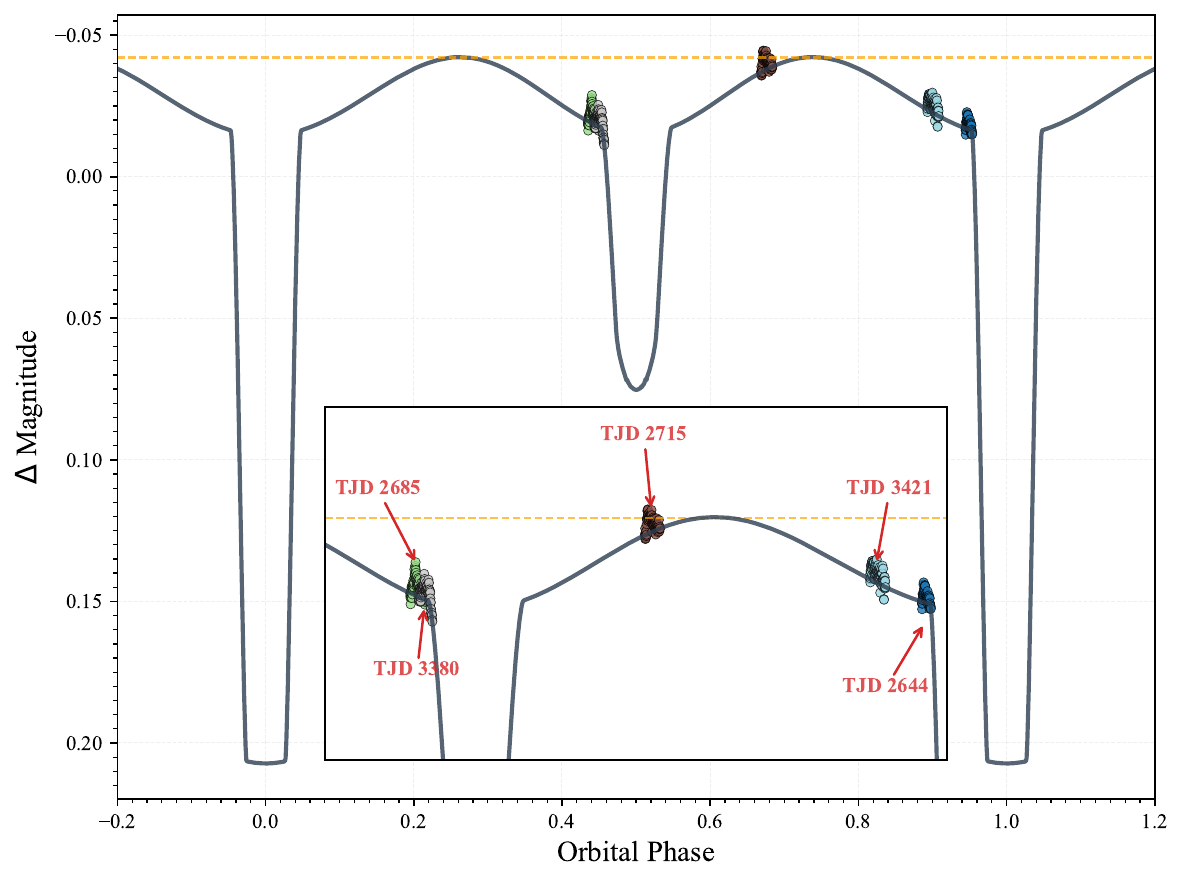}

	\caption{Flare events detected in TIC 207398432. The gray solid curve represents the modeled light curve of the eclipsing binary, while colored symbols mark the detected flare events. The inset shows a zoomed view of the flares, with arrows indicating their occurrence times in TJD.}
	\label{Fig.9}
\end{figure*}

On the other hand, flare activity differs significantly among the targets. In the LC of TIC 207398432, we detected five flare events occurring at TJD (= BJD - 2,457,000) 2644, 2685, 2715, 3380, and 3421. Using the same method as \citet{2021AJ....161...46S} to estimate flare energies, we found that these five flares have energies in the range of $6.97\times10^{35}$–$2.20\times10^{36} \mathrm{erg}$, all of which qualify as superflares. The properties of these flares are listed in Table \ref{tab:flares} and illustrated in Figure \ref{Fig.9}. No significant flare signals were detected in the other three targets, despite their similar masses. Therefore, the observed differences in flare activity are more likely related to variations in rotational state, magnetic field configuration, evolutionary stage, or to flares that are too weak to be detected.

Taken together, TIC 207398432 and KIC 10593759 exhibit stronger magnetic activity among the four systems, albeit in different aspects. TIC 207398432 shows the largest amplitude of spot-induced asymmetry and hosts multiple superflares, indicating intense magnetic activity. KIC 10593759, while exhibiting a similar level of asymmetry amplitude to the other two systems, shows the fastest temporal evolution of spot-induced asymmetry, reflecting rapidly evolving magnetic regions. The remaining two systems display lower asymmetry amplitudes and no significant flare events, suggesting weaker magnetic activity. This enhanced activity in TIC 207398432 and KIC 10593759 appears related to their evolutionary stages: components that have evolved into the subgiant or red-giant phases possess deeper convective envelopes, which, combined with still relatively rapid rotation in these close binaries, sustain efficient magnetic field generation. This is consistent with observations of active RS CVn-type binaries and highlights the value of twin binaries as benchmarks for stellar dynamo studies.

\begin{deluxetable*}{ccccc}
	\tablecaption{Properties of the Flares in TIC 207398432 \label{tab:flares}}
	\tablewidth{0pt}
	\tablehead{
		\colhead{Flare} & 
		\colhead{$t_\mathrm{begin}$ } & 
		\colhead{$t_\mathrm{end}$ } & 
		\colhead{$\Delta \mathrm{mag}$} & 
		\colhead{$E_\mathrm{flare}$ }\\
		\colhead{} & 
		\colhead{(TJD)} & 
		\colhead{(TJD)} & 
		\colhead{(mmag)} & 
		\colhead{(erg)}
	}
	\startdata
	1 & 2644.22698 & 2644.31032 & 6 & $6.97\times10^{35}$ \\
	2 & 2685.79685 & 2685.95658 & 10 & $1.91\times10^{36}$ \\
	3 & 2715.73415 & 2715.87303 & 7 & $1.85\times10^{36}$ \\
	4 & 3380.19761 & 3380.32261 & 8 & $1.64\times10^{36}$ \\
	5 & 3421.38914 & 3421.53081 & 7 & $2.20\times10^{36}$ \\
	\enddata
\end{deluxetable*}

%% Please use the acknowledgment and contribution environments. This will 
%% be anonomyized when the "anonymous" style option is used. 
\begin{acknowledgments}
This work is supported by the Yunnan Fundamental Research Projects (grant No.202503AP140013, 202501AS070055, 202401AS070046), the International Partnership Program of Chinese Academy of Sciences (No.020GJHZ2023030GC), the Chinese Academy of Sciences President's International Fellowship lnitiative (Grant No.2025PA0089), and the Yunnan Revitalization Talent Support Program. We also acknowledge the support of the staff at the 2.4-meter telescope of the Thai National Observatory (TNO) and the Guoshoujing Telescope (LAMOST, the Large Sky Area Multi-Object Fiber Spectroscopic Telescope) for their assistance in obtaining the spectroscopic observations. In addition, this research has made use of photometric data from the Kepler and TESS missions, accessed through the MAST data archive at the Space Telescope Science Institute (STScI). Funding for the TESS and Kepler missions is provided by the NASA Explorer Program and the NASA Science Mission Directorate, respectively. We further make use of spectroscopic data from the APOGEE survey, which is part of the Sloan Digital Sky Survey (SDSS).

\end{acknowledgments}

%% Similar to \facility{}, there is the optional \software command to allow 
%% authors a place to specify which programs were used during the creation of 
%% the manuscript. Authors should list each code and include either a
%% citation or url to the code inside ()s when available.

%% Appendix material should be preceded with a single \appendix command.
%% There should be a \section command for each appendix. Mark appendix
%% subsections with the same markup you use in the main body of the paper.
%%
%% Each Appendix (indicated with \section) will be lettered A, B, C, etc.
%% The equation counter will reset when it encounters the \appendix
%% command and will number appendix equations (A1), (A2), etc. The
%% Figure and Table counter will not reset.

\appendix

\renewcommand{\thetable}{A\arabic{table}}
\setcounter{table}{0}

\section{Radial Velocity Table}

\begin{deluxetable}{cccccc}
	\tablewidth{0pt}
	\tablecaption{RV Measurements of the Four Systems\label{tab:rv_long}}
	\tablehead{
		\colhead{Star} & \colhead{JD(Hel.)} & \colhead{Phase} & \colhead{RV$_1$ (km s$^{-1}$)} & \colhead{RV$_2$ (km s$^{-1}$)} & \colhead{Source}
	}
	\startdata
	\multirow{7}{*}{KIC 8957954}	&	2456931.70457 	&	0.565 	&	$	9.25 	\pm	0.86 	$	&	$	-58.89 	\pm	0.74 	$	&	SDSS	\\
	&	2457264.82207 	&	0.971 	&	$	-8.18 	\pm	1.13 	$	&	$	-41.50 	\pm	1.05 	$	&	SDSS	\\
	&	2457293.71953 	&	0.599 	&	$	24.67 	\pm	0.93 	$	&	$	-74.53 	\pm	1.04 	$	&	SDSS	\\
	&	\multicolumn{5}{c}{...}																	\\
	&	2457319.58329 	&	0.531 	&	$	-7.44 	\pm	0.85 	$	&	$	-40.44 	\pm	0.87 	$	&	SDSS	\\
	&	2457672.60145 	&	0.501 	&	$	-24.05 	\pm	0.85 	$	&	$	\ldots			$	&	SDSS	\\
	&	2458247.96356 	&	0.469 	&	$	-41.05 	\pm	0.86 	$	&	$	-8.44 	\pm	0.79 	$	&	SDSS	\\
	\hline																			
	\multirow{7}{*}{KIC 10593759}	&	2458181.99544 	&	0.329 	&	$	-118.00 	\pm	0.79 	$	&	$	15.72 	\pm	1.01 	$	&	SDSS	\\
	&	2458206.95510 	&	0.313 	&	$	-120.90 	\pm	0.86 	$	&	$	20.00 	\pm	0.82 	$	&	SDSS	\\
	&	2458234.87661 	&	0.770 	&	$	23.98 	\pm	1.08 	$	&	$	-127.99 	\pm	1.05 	$	&	SDSS	\\
	&	\multicolumn{5}{c}{...}																	\\
	&	2457942.71038 	&	0.137 	&	$	-109.21 	\pm	0.80 	$	&	$	7.16 	\pm	0.88 	$	&	SDSS	\\
	&	2458241.94774 	&	0.898 	&	$	-5.71 	\pm	0.96 	$	&	$	-98.13 	\pm	0.94 	$	&	SDSS	\\
	&	2458293.76536 	&	0.169 	&	$	-118.43 	\pm	0.97 	$	&	$	16.46 	\pm	0.86 	$	&	SDSS	\\
	\hline																			
	\multirow{7}{*}{KIC 8302455}	&	2458025.00351 	&	0.512 	&	$	\ldots			$	&	$	-12.61 	\pm	1.63 	$	&	LAMOST	\\
	&	2458025.01323 	&	0.514 	&	$	\ldots			$	&	$	-12.86 	\pm	2.06 	$	&	LAMOST	\\
	&	2458025.02226 	&	0.516 	&	$	\ldots			$	&	$	-12.63 	\pm	1.59 	$	&	LAMOST	\\
	&	\multicolumn{5}{c}{...}																	\\
	&	2459369.28123 	&	0.755 	&	$	73.52 	\pm	1.08 	$	&	$	-85.24 	\pm	1.31 	$	&	LAMOST	\\
	&	2459369.29790 	&	0.758 	&	$	73.41 	\pm	1.22 	$	&	$	-85.03 	\pm	1.19 	$	&	LAMOST	\\
	&	2459369.31387 	&	0.761 	&	$	73.24 	\pm	1.30 	$	&	$	-85.02 	\pm	1.18 	$	&	LAMOST	\\
	\hline																			
	\multirow{7}{*}{TIC 207398432}	&	2458246.25457 	&	0.937 	&	$	-20.54 	\pm	1.08 	$	&	$	-70.78 	\pm	2.07 	$	&	LAMOST	\\
	&	2458246.27054 	&	0.939 	&	$	-18.65 	\pm	1.13 	$	&	$	-69.03 	\pm	1.22 	$	&	LAMOST	\\
	&	2458246.28652 	&	0.940 	&	$	-25.38 	\pm	1.66 	$	&	$	-70.29 	\pm	1.88 	$	&	LAMOST	\\
	&	\multicolumn{5}{c}{...}																	\\
	&	2460740.40450 	&	0.353 	&	$	-95.17 	\pm	0.91 	$	&	$	12.25 	\pm	1.62 	$	&	TNO	\\
	&	2460757.24475 	&	0.172 	&	$	-103.30 	\pm	0.76 	$	&	$	16.34 	\pm	3.06 	$	&	TNO	\\
	&	2460757.34121 	&	0.182 	&	$	-103.35 	\pm	0.77 	$	&	$	17.30 	\pm	5.76 	$	&	TNO	\\
	\enddata
	\tablecomments{This table is available in its entirety in machine-readable form in the online article.}
\end{deluxetable}

%% For this sample we use BibTeX plus aasjournalv7.bst to generate the
%% the bibliography. The sample7.bib file was populated from ADS. To
%% get the citations to show in the compiled file do the following:
%%
%% pdflatex sample7.tex
%% bibtext sample7
%% pdflatex sample7.tex
%% pdflatex sample7.tex

\bibliography{sample7}{}

\begin{thebibliography}{}
\expandafter\ifx\csname natexlab\endcsname\relax\def\natexlab#1{#1}\fi
\providecommand{\url}[1]{\href{#1}{#1}}
\providecommand{\dodoi}[1]{doi:~\href{http://doi.org/#1}{\nolinkurl{#1}}}
\providecommand{\doeprint}[1]{\href{http://ascl.net/#1}{\nolinkurl{http://ascl.net/#1}}}
\providecommand{\doarXiv}[1]{\href{https://arxiv.org/abs/#1}{\nolinkurl{https://arxiv.org/abs/#1}}}

\bibitem[{M.~R. {Bate}(2000){Bate}}]{2000MNRAS.314...33B}
{Bate}, M.~R. 2000, \bibinfo{title}{{Predicting the properties of binary
  stellar systems: the evolution of accreting protobinary systems},} \mnras,
  314, 33, \dodoi{10.1046/j.1365-8711.2000.03333.x}

\bibitem[{M.~R. {Bate} {et~al.}(2002){Bate}, {Bonnell}, \&
  {Bromm}}]{2002MNRAS.336..705B}
{Bate}, M.~R., {Bonnell}, I.~A., \& {Bromm}, V. 2002, \bibinfo{title}{{The
  formation of close binary systems by dynamical interactions and orbital
  decay},} \mnras, 336, 705, \dodoi{10.1046/j.1365-8711.2002.05775.x}

\bibitem[{K. {Biazzo} {et~al.}(2006){Biazzo}, {Frasca}, {Catalano}, \&
  {Marilli}}]{2006A&A...446.1129B}
{Biazzo}, K., {Frasca}, A., {Catalano}, S., \& {Marilli}, E. 2006,
  \bibinfo{title}{{Photospheric and chromospheric active regions on three
  single-lined RS CVn binaries},} \aap, 446, 1129,
  \dodoi{10.1051/0004-6361:20053213}

\bibitem[{S. {Blanco-Cuaresma}(2019){Blanco-Cuaresma}}]{2019MNRAS.486.2075B}
{Blanco-Cuaresma}, S. 2019, \bibinfo{title}{{Modern stellar spectroscopy
  caveats},} \mnras, 486, 2075, \dodoi{10.1093/mnras/stz549}

\bibitem[{S. {Blanco-Cuaresma} {et~al.}(2014){Blanco-Cuaresma}, {Soubiran},
  {Heiter}, \& {Jofr{\'e}}}]{2014A&A...569A.111B}
{Blanco-Cuaresma}, S., {Soubiran}, C., {Heiter}, U., \& {Jofr{\'e}}, P. 2014,
  \bibinfo{title}{{Determining stellar atmospheric parameters and chemical
  abundances of FGK stars with iSpec},} \aap, 569, A111,
  \dodoi{10.1051/0004-6361/201423945}

\bibitem[{T. {Borkovits} {et~al.}(2016){Borkovits}, {Hajdu}, {Sztakovics},
  {Rappaport}, {Levine}, {B{\'\i}r{\'o}}, \& {Klagyivik}}]{2016MNRAS.455.4136B}
{Borkovits}, T., {Hajdu}, T., {Sztakovics}, J., {et~al.} 2016,
  \bibinfo{title}{{A comprehensive study of the Kepler triples via eclipse
  timing},} \mnras, 455, 4136, \dodoi{10.1093/mnras/stv2530}

\bibitem[{W.~J. {Borucki} {et~al.}(2010){Borucki}, {Koch}, {Basri}, {Batalha},
  {Brown}, {Caldwell}, {Caldwell}, {Christensen-Dalsgaard}, {Cochran},
  {DeVore}, {Dunham}, {Dupree}, {Gautier}, {Geary}, {Gilliland}, {Gould},
  {Howell}, {Jenkins}, {Kondo}, {Latham}, {Marcy}, {Meibom}, {Kjeldsen},
  {Lissauer}, {Monet}, {Morrison}, {Sasselov}, {Tarter}, {Boss}, {Brownlee},
  {Owen}, {Buzasi}, {Charbonneau}, {Doyle}, {Fortney}, {Ford}, {Holman},
  {Seager}, {Steffen}, {Welsh}, {Rowe}, {Anderson}, {Buchhave}, {Ciardi},
  {Walkowicz}, {Sherry}, {Horch}, {Isaacson}, {Everett}, {Fischer}, {Torres},
  {Johnson}, {Endl}, {MacQueen}, {Bryson}, {Dotson}, {Haas}, {Kolodziejczak},
  {Van Cleve}, {Chandrasekaran}, {Twicken}, {Quintana}, {Clarke}, {Allen},
  {Li}, {Wu}, {Tenenbaum}, {Verner}, {Bruhweiler}, {Barnes}, \&
  {Prsa}}]{2010Sci...327..977B}
{Borucki}, W.~J., {Koch}, D., {Basri}, G., {et~al.} 2010,
  \bibinfo{title}{{Kepler Planet-Detection Mission: Introduction and First
  Results},} Science, 327, 977, \dodoi{10.1126/science.1185402}

\bibitem[{A.~G. {Cantrell} \& T.~J. {Dougan}(2014){Cantrell} \&
  {Dougan}}]{2014MNRAS.445.2028C}
{Cantrell}, A.~G., \& {Dougan}, T.~J. 2014, \bibinfo{title}{{Twins like to be
  seen: observational biases affecting spectroscopically selected binary
  stars},} \mnras, 445, 2028, \dodoi{10.1093/mnras/stu1890}

\bibitem[{J. {Choi} {et~al.}(2016){Choi}, {Dotter}, {Conroy}, {Cantiello},
  {Paxton}, \& {Johnson}}]{2016ApJ...823..102C}
{Choi}, J., {Dotter}, A., {Conroy}, C., {et~al.} 2016, \bibinfo{title}{{Mesa
  Isochrones and Stellar Tracks (MIST). I. Solar-scaled Models},} \apj, 823,
  102, \dodoi{10.3847/0004-637X/823/2/102}

\bibitem[{X.-Q. {Cui} {et~al.}(2012){Cui}, {Zhao}, {Chu}, {Li}, {Li}, {Zhang},
  {Su}, {Yao}, {Wang}, {Xing}, {Li}, {Zhu}, {Wang}, {Gu}, {Luo}, {Xu}, {Zhang},
  {Liu}, {Zhang}, {Yang}, {Cao}, {Chen}, {Chen}, {Chen}, {Chen}, {Chu}, {Feng},
  {Gong}, {Hou}, {Hu}, {Hu}, {Hu}, {Jia}, {Jiang}, {Jiang}, {Jiang}, {Jin},
  {Li}, {Li}, {Li}, {Liu}, {Liu}, {Lu}, {Mao}, {Men}, {Qi}, {Qi}, {Shi},
  {Tang}, {Tao}, {Wang}, {Wang}, {Wang}, {Wang}, {Wang}, {Wang}, {Wang},
  {Wang}, {Wang}, {Wang}, {Wang}, {Wang}, {Xu}, {Xu}, {Yang}, {Yu}, {Yuan},
  {Yuan}, {Zhai}, {Zhang}, {Zhang}, {Zhang}, {Zhao}, {Zhou}, {Zhou}, {Zhu}, \&
  {Zou}}]{2012RAA....12.1197C}
{Cui}, X.-Q., {Zhao}, Y.-H., {Chu}, Y.-Q., {et~al.} 2012, \bibinfo{title}{{The
  Large Sky Area Multi-Object Fiber Spectroscopic Telescope (LAMOST)},}
  Research in Astronomy and Astrophysics, 12, 1197,
  \dodoi{10.1088/1674-4527/12/9/003}

\bibitem[{A. {Dotter}(2016){Dotter}}]{2016ApJS..222....8D}
{Dotter}, A. 2016, \bibinfo{title}{{MESA Isochrones and Stellar Tracks (MIST)
  0: Methods for the Construction of Stellar Isochrones},} \apjs, 222, 8,
  \dodoi{10.3847/0067-0049/222/1/8}

\bibitem[{P. {Eggenberger} {et~al.}(2010){Eggenberger}, {Meynet}, {Maeder},
  {Miglio}, {Montalban}, {Carrier}, {Mathis}, {Charbonnel}, \&
  {Talon}}]{2010A&A...519A.116E}
{Eggenberger}, P., {Meynet}, G., {Maeder}, A., {et~al.} 2010,
  \bibinfo{title}{{Effects of rotational mixing on the asteroseismic properties
  of solar-type stars},} \aap, 519, A116, \dodoi{10.1051/0004-6361/201014713}

\bibitem[{L. {Fellay} {et~al.}(2021){Fellay}, {Buldgen}, {Eggenberger}, {Khan},
  {Salmon}, {Miglio}, \& {Montalb{\'a}n}}]{2021A&A...654A.133F}
{Fellay}, L., {Buldgen}, G., {Eggenberger}, P., {et~al.} 2021,
  \bibinfo{title}{{Asteroseismology of evolved stars to constrain the internal
  transport of angular momentum. IV. Internal rotation of Kepler-56 from an
  MCMC analysis of the rotational splittings},} \aap, 654, A133,
  \dodoi{10.1051/0004-6361/202140518}

\bibitem[{D. {Garc{\'\i}a-Alvarez} {et~al.}(2003){Garc{\'\i}a-Alvarez},
  {Foing}, {Montes}, {Oliveira}, {Doyle}, {Messina}, {Lanza}, {Rodon{\`o}},
  {Abbott}, {Ash}, {Baldry}, {Bedding}, {Buckley}, {Cami}, {Cao}, {Catala},
  {Cheng}, {Domiciano de Souza}, {Donati}, {Hubert}, {Janot-Pacheco}, {Hao},
  {Kaper}, {Kaufer}, {Leister}, {Neff}, {Neiner}, {Orlando}, {O'Toole},
  {Sch{\"a}fer}, {Smartt}, {Stahl}, {Telting}, \&
  {Tubbesing}}]{2003A&A...397..285G}
{Garc{\'\i}a-Alvarez}, D., {Foing}, B.~H., {Montes}, D., {et~al.} 2003,
  \bibinfo{title}{{Simultaneous optical and X-ray observations of flares and
  rotational modulation on the RS CVn binary HR 1099 (V711 Tau) from the
  MUSICOS 1998 campaign},} \aap, 397, 285, \dodoi{10.1051/0004-6361:20021481}

\bibitem[{R.~L. {Gilliland} {et~al.}(2010){Gilliland}, {Jenkins}, {Borucki},
  {Bryson}, {Caldwell}, {Clarke}, {Dotson}, {Haas}, {Hall}, {Klaus}, {Koch},
  {McCauliff}, {Quintana}, {Twicken}, \& {van Cleve}}]{2010ApJ...713L.160G}
{Gilliland}, R.~L., {Jenkins}, J.~M., {Borucki}, W.~J., {et~al.} 2010,
  \bibinfo{title}{{Initial Characteristics of Kepler Short Cadence Data},}
  \apjl, 713, L160, \dodoi{10.1088/2041-8205/713/2/L160}

\bibitem[{H. {Hensberge} {et~al.}(2008){Hensberge}, {Iliji{\'c}}, \&
  {Torres}}]{2008A&A...482.1031H}
{Hensberge}, H., {Iliji{\'c}}, S., \& {Torres}, K.~B.~V. 2008,
  \bibinfo{title}{{On the separation of component spectra in binary and
  higher-multiplicity stellar systems: bias progression and spurious
  patterns},} \aap, 482, 1031, \dodoi{10.1051/0004-6361:20079038}

\bibitem[{J. {Higl} \& A. {Weiss}(2017){Higl} \& {Weiss}}]{2017A&A...608A..62H}
{Higl}, J., \& {Weiss}, A. 2017, \bibinfo{title}{{Testing stellar evolution
  models with detached eclipsing binaries},} \aap, 608, A62,
  \dodoi{10.1051/0004-6361/201731008}

\bibitem[{T.~O. {Husser} {et~al.}(2013){Husser}, {Wende-von Berg}, {Dreizler},
  {Homeier}, {Reiners}, {Barman}, \& {Hauschildt}}]{2013A&A...553A...6H}
{Husser}, T.~O., {Wende-von Berg}, S., {Dreizler}, S., {et~al.} 2013,
  \bibinfo{title}{{A new extensive library of PHOENIX stellar atmospheres and
  synthetic spectra},} \aap, 553, A6, \dodoi{10.1051/0004-6361/201219058}

\bibitem[{S. {Ilijic} {et~al.}(2004){Ilijic}, {Hensberge}, {Pavlovski}, \&
  {Freyhammer}}]{2004ASPC..318..111I}
{Ilijic}, S., {Hensberge}, H., {Pavlovski}, K., \& {Freyhammer}, L.~M. 2004, in
  Astronomical Society of the Pacific Conference Series, Vol. 318,
  Spectroscopically and Spatially Resolving the Components of the Close Binary
  Stars, ed. R.~W. {Hilditch}, H.~{Hensberge}, \& K.~{Pavlovski}, 111--113

\bibitem[{B. {Kirk} {et~al.}(2016){Kirk}, {Conroy}, {Pr{\v{s}}a},
  {Abdul-Masih}, {Kochoska}, {Matijevi{\v{c}}}, {Hambleton}, {Barclay},
  {Bloemen}, {Boyajian}, {Doyle}, {Fulton}, {Hoekstra}, {Jek}, {Kane},
  {Kostov}, {Latham}, {Mazeh}, {Orosz}, {Pepper}, {Quarles}, {Ragozzine},
  {Shporer}, {Southworth}, {Stassun}, {Thompson}, {Welsh}, {Agol}, {Derekas},
  {Devor}, {Fischer}, {Green}, {Gropp}, {Jacobs}, {Johnston}, {LaCourse},
  {Saetre}, {Schwengeler}, {Toczyski}, {Werner}, {Garrett}, {Gore}, {Martinez},
  {Spitzer}, {Stevick}, {Thomadis}, {Vrijmoet}, {Yenawine}, {Batalha}, \&
  {Borucki}}]{2016AJ....151...68K}
{Kirk}, B., {Conroy}, K., {Pr{\v{s}}a}, A., {et~al.} 2016,
  \bibinfo{title}{{Kepler Eclipsing Binary Stars. VII. The Catalog of Eclipsing
  Binaries Found in the Entire Kepler Data Set},} \aj, 151, 68,
  \dodoi{10.3847/0004-6256/151/3/68}

\bibitem[{M.~F. {Knote} {et~al.}(2022){Knote}, {Caballero-Nieves}, {Gokhale},
  {Johnston}, \& {Perlman}}]{2022ApJS..262...10K}
{Knote}, M.~F., {Caballero-Nieves}, S.~M., {Gokhale}, V., {Johnston}, K.~B., \&
  {Perlman}, E.~S. 2022, \bibinfo{title}{{Characteristics of Kepler Eclipsing
  Binaries Displaying a Significant O'Connell Effect},} \apjs, 262, 10,
  \dodoi{10.3847/1538-4365/ac770f}

\bibitem[{D.~G. {Koch} {et~al.}(2010){Koch}, {Borucki}, {Basri}, {Batalha},
  {Brown}, {Caldwell}, {Christensen-Dalsgaard}, {Cochran}, {DeVore}, {Dunham},
  {Gautier}, {Geary}, {Gilliland}, {Gould}, {Jenkins}, {Kondo}, {Latham},
  {Lissauer}, {Marcy}, {Monet}, {Sasselov}, {Boss}, {Brownlee}, {Caldwell},
  {Dupree}, {Howell}, {Kjeldsen}, {Meibom}, {Morrison}, {Owen}, {Reitsema},
  {Tarter}, {Bryson}, {Dotson}, {Gazis}, {Haas}, {Kolodziejczak}, {Rowe}, {Van
  Cleve}, {Allen}, {Chandrasekaran}, {Clarke}, {Li}, {Quintana}, {Tenenbaum},
  {Twicken}, \& {Wu}}]{2010ApJ...713L..79K}
{Koch}, D.~G., {Borucki}, W.~J., {Basri}, G., {et~al.} 2010,
  \bibinfo{title}{{Kepler Mission Design, Realized Photometric Performance, and
  Early Science},} \apjl, 713, L79, \dodoi{10.1088/2041-8205/713/2/L79}

\bibitem[{M. {Koleva} {et~al.}(2009){Koleva}, {Prugniel}, {Bouchard}, \&
  {Wu}}]{2009A&A...501.1269K}
{Koleva}, M., {Prugniel}, P., {Bouchard}, A., \& {Wu}, Y. 2009,
  \bibinfo{title}{{ULySS: a full spectrum fitting package},} \aap, 501, 1269,
  \dodoi{10.1051/0004-6361/200811467}

\bibitem[{E. {Lastennet} \& D. {Valls-Gabaud}(2002){Lastennet} \&
  {Valls-Gabaud}}]{2002A&A...396..551L}
{Lastennet}, E., \& {Valls-Gabaud}, D. 2002, \bibinfo{title}{{Detached
  double-lined eclipsing binaries as critical tests of stellar evolution. Age
  and metallicity determinations from the HR diagram},} \aap, 396, 551,
  \dodoi{10.1051/0004-6361:20021312}

\bibitem[{Y. {Lei} {et~al.}(2022){Lei}, {Li}, {Zhou}, \&
  {Li}}]{2022AJ....163..235L}
{Lei}, Y., {Li}, G., {Zhou}, G., \& {Li}, C. 2022, \bibinfo{title}{{Analysis of
  Five Double-lined Spectroscopic Eclipsing Binaries Observed with TESS and
  LAMOST},} \aj, 163, 235, \dodoi{10.3847/1538-3881/ac5aa5}

\bibitem[{J. {Liu} \& J. {Wu}(2024){Liu} \& {Wu}}]{2024ApJ...965..167L}
{Liu}, J., \& {Wu}, J. 2024, \bibinfo{title}{{The X-Ray Emission Reveals the
  Coronal Activities of Semi-detached Binaries},} \apj, 965, 167,
  \dodoi{10.3847/1538-4357/ad267e}

\bibitem[{L.~B. {Lucy}(2006){Lucy}}]{2006A&A...457..629L}
{Lucy}, L.~B. 2006, \bibinfo{title}{{Spectroscopic binaries with components of
  similar mass},} \aap, 457, 629, \dodoi{10.1051/0004-6361:20065746}

\bibitem[{L.~B. {Lucy} \& E. {Ricco}(1979){Lucy} \&
  {Ricco}}]{1979AJ.....84..401L}
{Lucy}, L.~B., \& {Ricco}, E. 1979, \bibinfo{title}{{The significance of
  binaries with nearly identical components.},} \aj, 84, 401,
  \dodoi{10.1086/112434}

\bibitem[{A.~L. {Luo} {et~al.}(2015){Luo}, {Zhao}, {Zhao}, {Deng}, {Liu},
  {Jing}, {Wang}, {Zhang}, {Shi}, {Cui}, {Chu}, {Li}, {Bai}, {Wu}, {Cai},
  {Cao}, {Cao}, {Carlin}, {Chen}, {Chen}, {Chen}, {Chen}, {Chen}, {Chen},
  {Chen}, {Christlieb}, {Chu}, {Cui}, {Dong}, {Du}, {Fan}, {Feng}, {Fu}, {Gao},
  {Gong}, {Gu}, {Guo}, {Han}, {He}, {Hou}, {Hou}, {Hou}, {Hu}, {Hu}, {Hu},
  {Huo}, {Jia}, {Jiang}, {Jiang}, {Jiang}, {Jin}, {Kong}, {Kong}, {Lei}, {Li},
  {Li}, {Li}, {Li}, {Li}, {Li}, {Li}, {Li}, {Li}, {Li}, {Li}, {Li}, {Liang},
  {Lin}, {Liu}, {Liu}, {Liu}, {Liu}, {Lu}, {Luo}, {Mao}, {Newberg}, {Ni}, {Qi},
  {Qi}, {Shen}, {Shi}, {Song}, {Song}, {Su}, {Su}, {Tang}, {Tao}, {Tian},
  {Wang}, {Wang}, {Wang}, {Wang}, {Wang}, {Wang}, {Wang}, {Wang}, {Wang},
  {Wang}, {Wang}, {Wang}, {Wang}, {Wang}, {Wang}, {Wang}, {Wang}, {Wang},
  {Wang}, {Wang}, {Wei}, {Wei}, {Wu}, {Wu}, {Wu}, {Wu}, {Xing}, {Xu}, {Xu},
  {Xu}, {Yan}, {Yang}, {Yang}, {Yang}, {Yang}, {Yao}, {Yu}, {Yuan}, {Yuan},
  {Yuan}, {Yuan}, {Zhai}, {Zhang}, {Zhang}, {Zhang}, {Zhang}, {Zhang}, {Zhang},
  {Zhang}, {Zhang}, {Zhao}, {Zhou}, {Zhou}, {Zhu}, {Zhu}, {Zou}, \&
  {Zuo}}]{2015RAA....15.1095L}
{Luo}, A.~L., {Zhao}, Y.-H., {Zhao}, G., {et~al.} 2015, \bibinfo{title}{{The
  first data release (DR1) of the LAMOST regular survey},} Research in
  Astronomy and Astrophysics, 15, 1095, \dodoi{10.1088/1674-4527/15/8/002}

\bibitem[{S.~R. {Majewski} {et~al.}(2017){Majewski}, {Schiavon}, {Frinchaboy},
  {Allende Prieto}, {Barkhouser}, {Bizyaev}, {Blank}, {Brunner}, {Burton},
  {Carrera}, {Chojnowski}, {Cunha}, {Epstein}, {Fitzgerald}, {Garc{\'\i}a
  P{\'e}rez}, {Hearty}, {Henderson}, {Holtzman}, {Johnson}, {Lam}, {Lawler},
  {Maseman}, {M{\'e}sz{\'a}ros}, {Nelson}, {Nguyen}, {Nidever}, {Pinsonneault},
  {Shetrone}, {Smee}, {Smith}, {Stolberg}, {Skrutskie}, {Walker}, {Wilson},
  {Zasowski}, {Anders}, {Basu}, {Beland}, {Blanton}, {Bovy}, {Brownstein},
  {Carlberg}, {Chaplin}, {Chiappini}, {Eisenstein}, {Elsworth}, {Feuillet},
  {Fleming}, {Galbraith-Frew}, {Garc{\'\i}a}, {Garc{\'\i}a-Hern{\'a}ndez},
  {Gillespie}, {Girardi}, {Gunn}, {Hasselquist}, {Hayden}, {Hekker}, {Ivans},
  {Kinemuchi}, {Klaene}, {Mahadevan}, {Mathur}, {Mosser}, {Muna}, {Munn},
  {Nichol}, {O'Connell}, {Parejko}, {Robin}, {Rocha-Pinto}, {Schultheis},
  {Serenelli}, {Shane}, {Silva Aguirre}, {Sobeck}, {Thompson}, {Troup},
  {Weinberg}, \& {Zamora}}]{2017AJ....154...94M}
{Majewski}, S.~R., {Schiavon}, R.~P., {Frinchaboy}, P.~M., {et~al.} 2017,
  \bibinfo{title}{{The Apache Point Observatory Galactic Evolution Experiment
  (APOGEE)},} \aj, 154, 94, \dodoi{10.3847/1538-3881/aa784d}

\bibitem[{P.~F.~L. {Maxted} {et~al.}(2020){Maxted}, {Gaulme}, {Graczyk},
  {He{\l}miniak}, {Johnston}, {Orosz}, {Pr{\v{s}}a}, {Southworth}, {Torres},
  {Davies}, {Ball}, \& {Chaplin}}]{2020MNRAS.498..332M}
{Maxted}, P.~F.~L., {Gaulme}, P., {Graczyk}, D., {et~al.} 2020,
  \bibinfo{title}{{The TESS light curve of AI Phoenicis},} \mnras, 498, 332,
  \dodoi{10.1093/mnras/staa1662}

\bibitem[{S.~A. {McCartney}(1999){McCartney}}]{1999PhDT........38M}
{McCartney}, S.~A. 1999, PhD thesis, University of Oklahoma, Norman

\bibitem[{F. {Meng} {et~al.}(2023){Meng}, {Zhu}, {Qian}, {Liu}, {Li}, \&
  {Matekov}}]{2023ApJ...954..111M}
{Meng}, F., {Zhu}, L., {Qian}, S., {et~al.} 2023, \bibinfo{title}{{NY Bootes:
  An Active Deep and Low-mass-ratio Contact Binary with a Cool Companion in a
  Hierarchical Triple System},} \apj, 954, 111,
  \dodoi{10.3847/1538-4357/ace8fe}

\bibitem[{D. {Montes} {et~al.}(1996){Montes}, {Fernandez-Figueroa}, {Cornide},
  \& {de Castro}}]{1996A&A...312..221M}
{Montes}, D., {Fernandez-Figueroa}, M.~J., {Cornide}, M., \& {de Castro}, E.
  1996, \bibinfo{title}{{The behaviour of the excess CA II H and K and Hɛ
  emissions in chromospherically active binaries.},} \aap, 312, 221,
  \dodoi{10.48550/arXiv.astro-ph/9511125}

\bibitem[{S. {Naoz}(2016){Naoz}}]{2016ARA&A..54..441N}
{Naoz}, S. 2016, \bibinfo{title}{{The Eccentric Kozai-Lidov Effect and Its
  Applications},} \araa, 54, 441, \dodoi{10.1146/annurev-astro-081915-023315}

\bibitem[{Y. {Pan} \& X. {Zhang}(2023){Pan} \& {Zhang}}]{2023AJ....165..247P}
{Pan}, Y., \& {Zhang}, X. 2023, \bibinfo{title}{{KIC 7284688: A Solar-type
  Eclipsing Binary with Rapidly Varying O'Connell Effect},} \aj, 165, 247,
  \dodoi{10.3847/1538-3881/accfa1}

\bibitem[{P. {Prugniel} \& C. {Soubiran}(2001){Prugniel} \&
  {Soubiran}}]{2001A&A...369.1048P}
{Prugniel}, P., \& {Soubiran}, C. 2001, \bibinfo{title}{{A database of high and
  medium-resolution stellar spectra},} \aap, 369, 1048,
  \dodoi{10.1051/0004-6361:20010163}

\bibitem[{P. {Prugniel} {et~al.}(2007){Prugniel}, {Soubiran}, {Koleva}, \& {Le
  Borgne}}]{2007astro.ph..3658P}
{Prugniel}, P., {Soubiran}, C., {Koleva}, M., \& {Le Borgne}, D. 2007,
  \bibinfo{title}{{New release of the ELODIE library: Version 3.1},} arXiv
  e-prints, astro, \dodoi{10.48550/arXiv.astro-ph/0703658}

\bibitem[{A. {Pr{\v{s}}a} {et~al.}(2022){Pr{\v{s}}a}, {Kochoska}, {Conroy},
  {Eisner}, {Hey}, {IJspeert}, {Kruse}, {Fleming}, {Johnston}, {Kristiansen},
  {LaCourse}, {Mortensen}, {Pepper}, {Stassun}, {Torres}, {Abdul-Masih},
  {Chakraborty}, {Gagliano}, {Guo}, {Hambleton}, {Hong}, {Jacobs}, {Jones},
  {Kostov}, {Lee}, {Omohundro}, {Orosz}, {Page}, {Powell}, {Rappaport}, {Reed},
  {Schnittman}, {Schwengeler}, {Shporer}, {Terentev}, {Vanderburg}, {Welsh},
  {Caldwell}, {Doty}, {Jenkins}, {Latham}, {Ricker}, {Seager}, {Schlieder},
  {Shiao}, {Vanderspek}, \& {Winn}}]{2022ApJS..258...16P}
{Pr{\v{s}}a}, A., {Kochoska}, A., {Conroy}, K.~E., {et~al.} 2022,
  \bibinfo{title}{{TESS Eclipsing Binary Stars. I. Short-cadence Observations
  of 4584 Eclipsing Binaries in Sectors 1-26},} \apjs, 258, 16,
  \dodoi{10.3847/1538-4365/ac324a}

\bibitem[{G.~R. {Ricker} {et~al.}(2015){Ricker}, {Winn}, {Vanderspek},
  {Latham}, {Bakos}, {Bean}, {Berta-Thompson}, {Brown}, {Buchhave}, {Butler},
  {Butler}, {Chaplin}, {Charbonneau}, {Christensen-Dalsgaard}, {Clampin},
  {Deming}, {Doty}, {De Lee}, {Dressing}, {Dunham}, {Endl}, {Fressin}, {Ge},
  {Henning}, {Holman}, {Howard}, {Ida}, {Jenkins}, {Jernigan}, {Johnson},
  {Kaltenegger}, {Kawai}, {Kjeldsen}, {Laughlin}, {Levine}, {Lin}, {Lissauer},
  {MacQueen}, {Marcy}, {McCullough}, {Morton}, {Narita}, {Paegert}, {Palle},
  {Pepe}, {Pepper}, {Quirrenbach}, {Rinehart}, {Sasselov}, {Sato}, {Seager},
  {Sozzetti}, {Stassun}, {Sullivan}, {Szentgyorgyi}, {Torres}, {Udry}, \&
  {Villasenor}}]{2015JATIS...1a4003R}
{Ricker}, G.~R., {Winn}, J.~N., {Vanderspek}, R., {et~al.} 2015,
  \bibinfo{title}{{Transiting Exoplanet Survey Satellite (TESS)},} Journal of
  Astronomical Telescopes, Instruments, and Systems, 1, 014003,
  \dodoi{10.1117/1.JATIS.1.1.014003}

\bibitem[{S. {Rucinski}(1999){Rucinski}}]{1999TJPh...23..271R}
{Rucinski}, S. 1999, \bibinfo{title}{{Spectral Broadening Functions},} Turkish
  Journal of Physics, 23, 271

\bibitem[{X.-d. {Shi} {et~al.}(2021){Shi}, {Qian}, {Li}, \&
  {Liu}}]{2021AJ....161...46S}
{Shi}, X.-d., {Qian}, S.-b., {Li}, L.-j., \& {Liu}, N.-p. 2021,
  \bibinfo{title}{{Flaring and Spot Activities on the Semi-detached Binary
  System KIC 06852488},} \aj, 161, 46, \dodoi{10.3847/1538-3881/abccd7}

\bibitem[{T. {Shibayama} {et~al.}(2013){Shibayama}, {Maehara}, {Notsu},
  {Notsu}, {Nagao}, {Honda}, {Ishii}, {Nogami}, \&
  {Shibata}}]{2013ApJS..209....5S}
{Shibayama}, T., {Maehara}, H., {Notsu}, S., {et~al.} 2013,
  \bibinfo{title}{{Superflares on Solar-type Stars Observed with Kepler. I.
  Statistical Properties of Superflares},} \apjs, 209, 5,
  \dodoi{10.1088/0067-0049/209/1/5}

\bibitem[{M. {Simon} \& R.~C. {Obbie}(2009){Simon} \&
  {Obbie}}]{2009AJ....137.3442S}
{Simon}, M., \& {Obbie}, R.~C. 2009, \bibinfo{title}{{Twins Among the Low-Mass
  Spectroscopic Binaries},} \aj, 137, 3442,
  \dodoi{10.1088/0004-6256/137/2/3442}

\bibitem[{K.~G. {Strassmeier}(2009){Strassmeier}}]{2009A&ARv..17..251S}
{Strassmeier}, K.~G. 2009, \bibinfo{title}{{Starspots},} \aapr, 17, 251,
  \dodoi{10.1007/s00159-009-0020-6}

\bibitem[{V. {Suntharalingam} {et~al.}(2023){Suntharalingam}, {Prigozhin},
  {Warner}, {Young}, {Woods}, \& {Berthiaume}}]{2023AN....34430139S}
{Suntharalingam}, V., {Prigozhin}, I., {Warner}, K., {et~al.} 2023,
  \bibinfo{title}{{CCD imagers for the Transiting Exoplanet Survey Satellite
  from benchtop to space environment},} Astronomische Nachrichten, 344,
  e20230139, \dodoi{10.1002/asna.20230139}

\bibitem[{A.~A. {Tokovinin}(2000){Tokovinin}}]{2000A&A...360..997T}
{Tokovinin}, A.~A. 2000, \bibinfo{title}{{On the origin of binaries with twin
  components},} \aap, 360, 997

\bibitem[{G. {Torres} {et~al.}(2010){Torres}, {Andersen}, \&
  {Gim{\'e}nez}}]{2010A&ARv..18...67T}
{Torres}, G., {Andersen}, J., \& {Gim{\'e}nez}, A. 2010,
  \bibinfo{title}{{Accurate masses and radii of normal stars: modern results
  and applications},} \aapr, 18, 67, \dodoi{10.1007/s00159-009-0025-1}

\bibitem[{K. {Tran} {et~al.}(2013){Tran}, {Levine}, {Rappaport}, {Borkovits},
  {Csizmadia}, \& {Kalomeni}}]{2013ApJ...774...81T}
{Tran}, K., {Levine}, A., {Rappaport}, S., {et~al.} 2013, \bibinfo{title}{{The
  Anticorrelated Nature of the Primary and Secondary Eclipse Timing Variations
  for the Kepler Contact Binaries},} \apj, 774, 81,
  \dodoi{10.1088/0004-637X/774/1/81}

\bibitem[{W. {van Hamme}(1993){van Hamme}}]{1993AJ....106.2096V}
{van Hamme}, W. 1993, \bibinfo{title}{{New Limb-Darkening Coefficients for
  Modeling Binary Star Light Curves},} \aj, 106, 2096, \dodoi{10.1086/116788}

\bibitem[{J. {Wang} {et~al.}(2024){Wang}, {Pan}, {Fu}, {Zong}, {Zong}, {Cang},
  {Zhang}, \& {Pan}}]{2024A&A...690A.201W}
{Wang}, J., {Pan}, Y., {Fu}, J., {et~al.} 2024, \bibinfo{title}{{Lifetime of
  starspots on detached eclipsing binaries: Detecting the effects of tides on
  stellar activity},} \aap, 690, A201, \dodoi{10.1051/0004-6361/202449484}

\bibitem[{R.~E. {Wilson} \& E.~J. {Devinney}(1971){Wilson} \&
  {Devinney}}]{1971ApJ...166..605W}
{Wilson}, R.~E., \& {Devinney}, E.~J. 1971, \bibinfo{title}{{Realization of
  Accurate Close-Binary Light Curves: Application to MR Cygni},} \apj, 166,
  605, \dodoi{10.1086/150986}

\bibitem[{R.~E. {Wilson} \& W. {Van Hamme}(2014){Wilson} \& {Van
  Hamme}}]{2014ApJ...780..151W}
{Wilson}, R.~E., \& {Van Hamme}, W. 2014, \bibinfo{title}{{Unification of
  Binary Star Ephemeris Solutions},} \apj, 780, 151,
  \dodoi{10.1088/0004-637X/780/2/151}

\bibitem[{Y. {Wu} {et~al.}(2014){Wu}, {Du}, {Luo}, {Zhao}, \&
  {Yuan}}]{2014IAUS..306..340W}
{Wu}, Y., {Du}, B., {Luo}, A., {Zhao}, Y., \& {Yuan}, H. 2014, in IAU
  Symposium, Vol. 306, Statistical Challenges in 21st Century Cosmology, ed.
  A.~{Heavens}, J.-L. {Starck}, \& A.~{Krone-Martins}, 340--342,
  \dodoi{10.1017/S1743921314010825}

\bibitem[{D.~G. {York} {et~al.}(2000){York}, {Adelman}, {Anderson}, {Anderson},
  {Annis}, {Bahcall}, {Bakken}, {Barkhouser}, {Bastian}, {Berman}, {Boroski},
  {Bracker}, {Briegel}, {Briggs}, {Brinkmann}, {Brunner}, {Burles}, {Carey},
  {Carr}, {Castander}, {Chen}, {Colestock}, {Connolly}, {Crocker}, {Csabai},
  {Czarapata}, {Davis}, {Doi}, {Dombeck}, {Eisenstein}, {Ellman}, {Elms},
  {Evans}, {Fan}, {Federwitz}, {Fiscelli}, {Friedman}, {Frieman}, {Fukugita},
  {Gillespie}, {Gunn}, {Gurbani}, {de Haas}, {Haldeman}, {Harris}, {Hayes},
  {Heckman}, {Hennessy}, {Hindsley}, {Holm}, {Holmgren}, {Huang}, {Hull},
  {Husby}, {Ichikawa}, {Ichikawa}, {Ivezi{\'c}}, {Kent}, {Kim}, {Kinney},
  {Klaene}, {Kleinman}, {Kleinman}, {Knapp}, {Korienek}, {Kron}, {Kunszt},
  {Lamb}, {Lee}, {Leger}, {Limmongkol}, {Lindenmeyer}, {Long}, {Loomis},
  {Loveday}, {Lucinio}, {Lupton}, {MacKinnon}, {Mannery}, {Mantsch}, {Margon},
  {McGehee}, {McKay}, {Meiksin}, {Merelli}, {Monet}, {Munn}, {Narayanan},
  {Nash}, {Neilsen}, {Neswold}, {Newberg}, {Nichol}, {Nicinski}, {Nonino},
  {Okada}, {Okamura}, {Ostriker}, {Owen}, {Pauls}, {Peoples}, {Peterson},
  {Petravick}, {Pier}, {Pope}, {Pordes}, {Prosapio}, {Rechenmacher}, {Quinn},
  {Richards}, {Richmond}, {Rivetta}, {Rockosi}, {Ruthmansdorfer}, {Sandford},
  {Schlegel}, {Schneider}, {Sekiguchi}, {Sergey}, {Shimasaku}, {Siegmund},
  {Smee}, {Smith}, {Snedden}, {Stone}, {Stoughton}, {Strauss}, {Stubbs},
  {SubbaRao}, {Szalay}, {Szapudi}, {Szokoly}, {Thakar}, {Tremonti}, {Tucker},
  {Uomoto}, {Vanden Berk}, {Vogeley}, {Waddell}, {Wang}, {Watanabe},
  {Weinberg}, {Yanny}, {Yasuda}, \& {SDSS Collaboration}}]{2000AJ....120.1579Y}
{York}, D.~G., {Adelman}, J., {Anderson}, Jr., J.~E., {et~al.} 2000,
  \bibinfo{title}{{The Sloan Digital Sky Survey: Technical Summary},} \aj, 120,
  1579, \dodoi{10.1086/301513}

\bibitem[{J. {Zhang} {et~al.}(2019){Zhang}, {Qian}, {Wu}, \&
  {Zhou}}]{2019ApJS..244...43Z}
{Zhang}, J., {Qian}, S.-B., {Wu}, Y., \& {Zhou}, X. 2019,
  \bibinfo{title}{{Unbiased Distribution of Binary Parameters from LAMOST and
  Kepler Observations},} \apjs, 244, 43, \dodoi{10.3847/1538-4365/ab442b}

\end{thebibliography}
\bibliographystyle{aasjournalv7}

%% This command is needed to show the entire author+affiliation list when
%% the collaboration and author truncation commands are used.  It has to
%% go at the end of the manuscript.
%\allauthors

%% Include this line if you are using the \added, \replaced, \deleted
%% commands to see a summary list of all changes at the end of the article.
%\listofchanges

\end{document}